# Ion exchange in atomically thin clays and micas

Yi-Chao Zou[1,2], Lucas Mogg[3,4,5], Nick Clark[2,3], Cihan Bacaksiz[6], Slavisa Milanovic[6], Vishnu Sreepal[3,7], Guang-Ping Hao[3,4], Yi-Chi Wang[2,8], David G. Hopkinson[2,3], Roman Gorbachev[3,4], Samuel Shaw[9], Kostya S. Novoselov[3,4], Rahul Raveendran-Nair[3,7], Francois M. Peeters[6], Marcelo Lozada-Hidalgo[3,4]*, Sarah J. Haigh[2,3]*

[1]School of Materials Science and Engineering, Sun Yat-sen University, Guangzhou, 510275, P. R. China

[2]Department of Materials, The University of Manchester, Manchester M13 9PL, UK

[3]National Graphene Institute, The University of Manchester, Manchester M13 9PL, UK

[4]Department of Physics and Astronomy, The University of Manchester, Manchester M13 9PL, UK

[5]Department of Engineering, University of Cambridge, 9 JJ Thomson Avenue, Cambridge CB3 0FA, UK

[6]Departement Fysica, Universiteit Antwerpen, Groenenborgerlaan 171, B-2020 Antwerp, Belgium

[7]Department of Chemical Engineering and Analytical Science, The University of Manchester, Manchester, M13 9PL, UK

[8]Beijing Institute of Nanoenergy and Nanosystems, Chinese Academy of Sciences, Beijing 101400, P. R. China

[9]Research Centre for Radwaste Disposal and Williamson Research Centre, School of Earth and Environmental Science, The University of Manchester, Manchester M13 9PL, UK

sarah.haigh@manchester.ac.uk
marcelo.lozadahidalgo@manchester.ac.uk

**Clays and micas are receiving attention as materials that, in their atomically thin form, could allow for novel proton conductive[1], ion selective[2-4], osmotic power generation[5], or solvent filtration membranes[6]. The interest arises from the possibility of controlling their properties by exchanging ions in the crystal lattice[1,6]. However, the ion exchange process itself remains largely unexplored in atomically thin materials. Here we use atomic-resolution scanning transmission electron microscopy to study the dynamics of the process and reveal the binding sites of individual ions in atomically thin and artificially restacked clays and micas. Imaging ion exchange after different exposure time and for different crystal thicknesses, we find that the ion diffusion constant, *D*, for the interlayer space of atomically thin samples is up to $10^4$ times larger than in bulk crystals and approaches its value in free water. Surprisingly, samples where no bulk exchange is expected display fast exchange if the mica layers are twisted and restacked; but in this case, the exchanged ions arrange in islands controlled by the moiré superlattice dimensions. We attribute the fast ion diffusion to enhanced interlayer expandability resulting from weaker interlayer binding forces in both atomically thin and restacked materials. Finally, we demonstrate images of individual surface cations for these materials, which had remained elusive in previous studies. This work provides atomic scale insights into ion diffusion in highly confined spaces and suggests strategies to design novel exfoliated clays membranes.**

Clays and micas are minerals that consist of aluminosilicate layers with cations both adsorbed on the basal plane surfaces and residing in the space between the layers. The native cations (typically $K^+$ and $Mg^{2+}$) can be exchanged for others when the material is exposed to electrolytes *via* a process known as ion exchange[7]. The crystals are relatively easy to exfoliate along the basal planes, producing high aspect ratio atomically thin sheets. Recent experiments demonstrated that exchanging the native cations for protons in atomically thin micas yields highly conductive proton transport membranes[1]. Exfoliated two-dimensional (2D) clay or mica crystallites can also be restacked to produce laminate membranes and composites, *via* similar fabrication methods as those used for graphene oxide[8,9]. The space between the restacked crystallites constitute channels with one of their dimensions comparable with, or narrower than, the Debye length in common electrolytes[3,4]; hence, water and ion transport through these channels depends critically on the crystal surface charge[10]. This allows for the use of clays and micas' ion exchange properties to influence phenomena in the channels, such as osmotic power generation[5], ion selective transport[2-4] or solvent filtration[6]. However, despite its importance to the different applications of these materials, ion exchange in atomically thin clays and micas remains largely unexplored. In this work, we used advanced aberration corrected scanning transmission electron microscopy (STEM) to study this phenomenon. STEM, unlike the more commonly used techniques such as atomic force microscopy or X-ray measurements[11-14], reveals the local binding environment of exchanged ions with sub-angstrom resolution, enabling understanding of the ion exchange process.

Atomic resolution STEM investigations of $Cs^+$ ion exchange were performed for atomically thin clays and micas. $Cs^+$ was chosen for this study because diffusion of radiocaesium ($^{137}Cs$) in clay soils is of environmental relevance following nuclear incidents[17], and because its large atomic number provides bright contrast in annular dark field (ADF)-STEM images. To characterise the ion exchange process, we obtained three-dimensional lattice images by combining both cross-sectional (Fig. 1a) and plan view STEM imaging (Fig. 1b). For cross-sectional imaging, the crystals were mechanically exfoliated onto oxidised silicon substrates and immersed in a $CsNO_3$ electrolyte (0.1 M aqueous solution) for a period from one second to several months to enable ion exchange. The crystals were then covered by a graphite flake and Pt protective strap, so that thin slices of this assembled stack could be extracted *via* focused ion beam milling[18] (Supplementary information, SI Section 2 and Fig. S1-S7). To produce plan view specimens, mechanically exfoliated mono- (1L), bi- (2L) and trilayer (3L) crystals were transferred over holes in $SiN_x$ TEM grids and immersed in $CsNO_3$ electrolyte, as described above. For further information on sample preparation, see Fig. S1-S4, and SI Section 2.1. In the following three sections, we describe our observations of ion exchange in the interlayer spaces of natural and twisted specimens as well as on the crystal basal plane surface. In each section, we exploited the differences in crystal chemistry[15,16] of one of three crystals – muscovite mica, biotite mica and vermiculite – to enable the highest possible quality images for each experiment (see SI Section 1, Tables S1-S2 and Section 8).

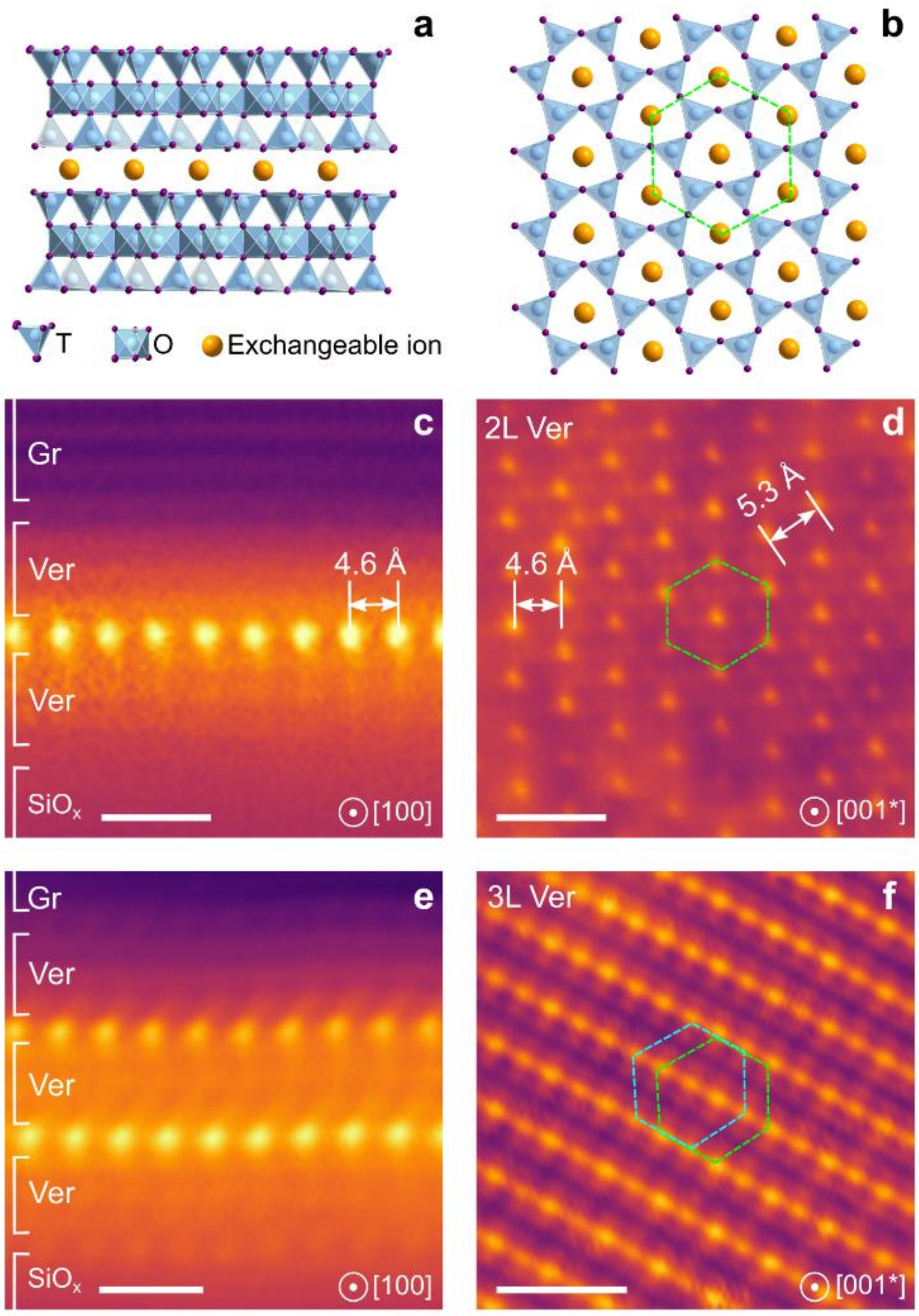

**Figure 1| Interlayer cations in exchanged clays and micas. a,** Cross-sectional atomic model of a typical pristine bilayer (2L) clay or mica viewed along the layers. The aluminosilicate tetrahedral-octahedral-tetrahedral, (TOT) layers are connected by exchangeable ions (e.g. $K^+$ and $Mg^{2+}$). The weakly bonded surface cations have been omitted for clarity. **b**, Plan view atomic model showing the interlayer ions with adjacent top and bottom T layers consisting of aluminosilicate (Al-Si-O) hexagonal rings. The exchangeable interlayer ions form a quasi-hexagonal lattice (marked with the green hexagon). **c**, **e** Cross-sectional annular dark field (ADF)-STEM images of Cs-exchanged **c,** bilayer and **e**, trilayer (3L) vermiculite (Ver). The interlayer ions are exchanged with $Cs^+$, visible as rows of bright dots in between the aluminosilicate layers. The atomically thin vermiculite is encapsulated between graphite (Gr) and $SiO_x$ to enable atomic resolution imaging. **d, f** Plan view ADF-STEM images of Cs-exchanged **d,** bilayer and **f,** trilayer vermiculite. A single interlayer of $Cs^+$ ions forms a quasi-hexagonal lattice (marked with the green hexagon in **d**), while the offset between two superimposed planes of interlayer $Cs^+$ ions creates a linear pattern (**f**). All scale bars, 1 nm.

## Ion exchange in the interlayer spaces of vermiculite clay

Fig. 1 shows ADF-STEM images taken from typical samples of Cs-exchanged atomically thin vermiculite – the fastest ion exchanger in this study. $Cs^+$ ions are visible as one and two rows of bright spots between the aluminosilicate layers in bilayer and trilayer cross-sectional specimens, respectively (Fig. 1c,e, Fig. S6,S7). These observations can be complemented with plan view imaging, which resolves the lateral distribution of the ions (Fig. 1d,f). In a trilayer sample, the projected arrangement of the two $Cs^+$ planes appears as a linear pattern formed by a superposition of two of these quasi-hexagonal $Cs^+$ lattices (Fig. 1f, and Fig. S9). These images confirm that the exchanged ions form fully occupied interlayer planes – one in 2L samples, two in 3L – with each layer having quasi-hexagonal symmetry (in-plane $Cs^+$ lattice constant of 5.3 Å). This uniform lateral distribution of $Cs^+$ suggests that the ion diffusion in the interlayer is not driven by in-plane defects like dislocations. In our samples, we see no preferential direction for ion penetration within the basal plane, indicating that ion diffusion is the result of uniform ion penetration into the interlayer space through the entire length of exposed crystal edges (see also SI Fig. S12).

To characterise the speed of the ion penetration into the interlayer space, we imaged >30 samples with thicknesses ranging from 2 to 70 layers that were subject to the $CsNO_3$ solution treatment for different lengths of time ($t$). These images effectively provide 'snapshots' of the exchange process. Cross-section images reveal the penetration distance of $Cs^+$ into the crystal, $\Delta P$ (Fig. 2), which allows for the estimation of the ion diffusion coefficient $D = \langle \Delta P^2 \rangle / 2t$ (SI Section 6)[19], with $\langle \Delta P^2 \rangle$ the mean square penetration distance averaged from measurements of all the individual interlayer spaces in the sample. Taking the 5L sample in Fig. 2b as an example ($t$ = 10 s), for each interlayer space (i1, i2, i3, i4), we measured the $Cs^+$ penetration distance $\Delta P$ ($\Delta P_1$, $\Delta P_2$, $\Delta P_3$, $\Delta P_4$), which allows for the estimation of the mean square displacement of $Cs^+$ as $\langle \Delta P^2 \rangle = \frac{\sum_{i=1}^{4} \Delta P_i^2}{4}$ = 3.0 µm$^2$, and the $Cs^+$ diffusivity $D_{5L\text{-}Ver} \approx 0.15$ µm$^2$s$^{-1}$. Following this method, we performed a systematic study of ion diffusion in vermiculite clays with thickness ranging from 2-70 layers. To our surprise, 2L samples showed $Cs^+$ exchange with large penetration distances of around 15 µm after only one second of $CsNO_3$ solution treatment (Fig. 2a, SI Section 6). This exceptionally fast exchange yields the ion exchange diffusion constant of bilayer vermiculite as ~$10^2$ µm$^2$ s$^{-1}$ – only ~10 times slower than the diffusion coefficient of $Cs^+$ in water[20]. This behaviour was reproducible across 3 different 2L samples and statistical analysis of these data revealed $D_{2L\text{-}Ver}$ = 60 ± 40 µm$^2$ s$^{-1}$. Similar measurements for 3L vermiculite showed a slower interlayer diffusivity value $D_{3L\text{-}Ver}$ = 15 ± 12 µm$^2$ s$^{-1}$. Repeating this experiment for different crystal thicknesses revealed that $D$ decreases exponentially with crystal thickness, reaching the literature values for $Cs^+$ diffusion in bulk vermiculite[16,21], $D_{Bulk\text{-}Ver}$ = (7±3) x $10^{-3}$ µm$^2$ s$^{-1}$ (Fig. 2c, d), for all specimens thicker than ~10 layers. The found dependence of $D$ on the number of layers, $N$, can be described by the formula $D = D_0\,[1+\exp(-\alpha(N-N_0))]$, where $\alpha$ = 1.17, a material-specific coefficient, $D_o$ = 5 x $10^{-3}$ µm$^2$ s$^{-1}$ and $N_0$ = 10. The fast exchange was confirmed using vermiculite laminates made from exfoliated thin flakes, which showed a similar enhancement of the ion-exchange speed compared to the bulk non-exfoliated crystal (SI Section 7, and Fig. S11).

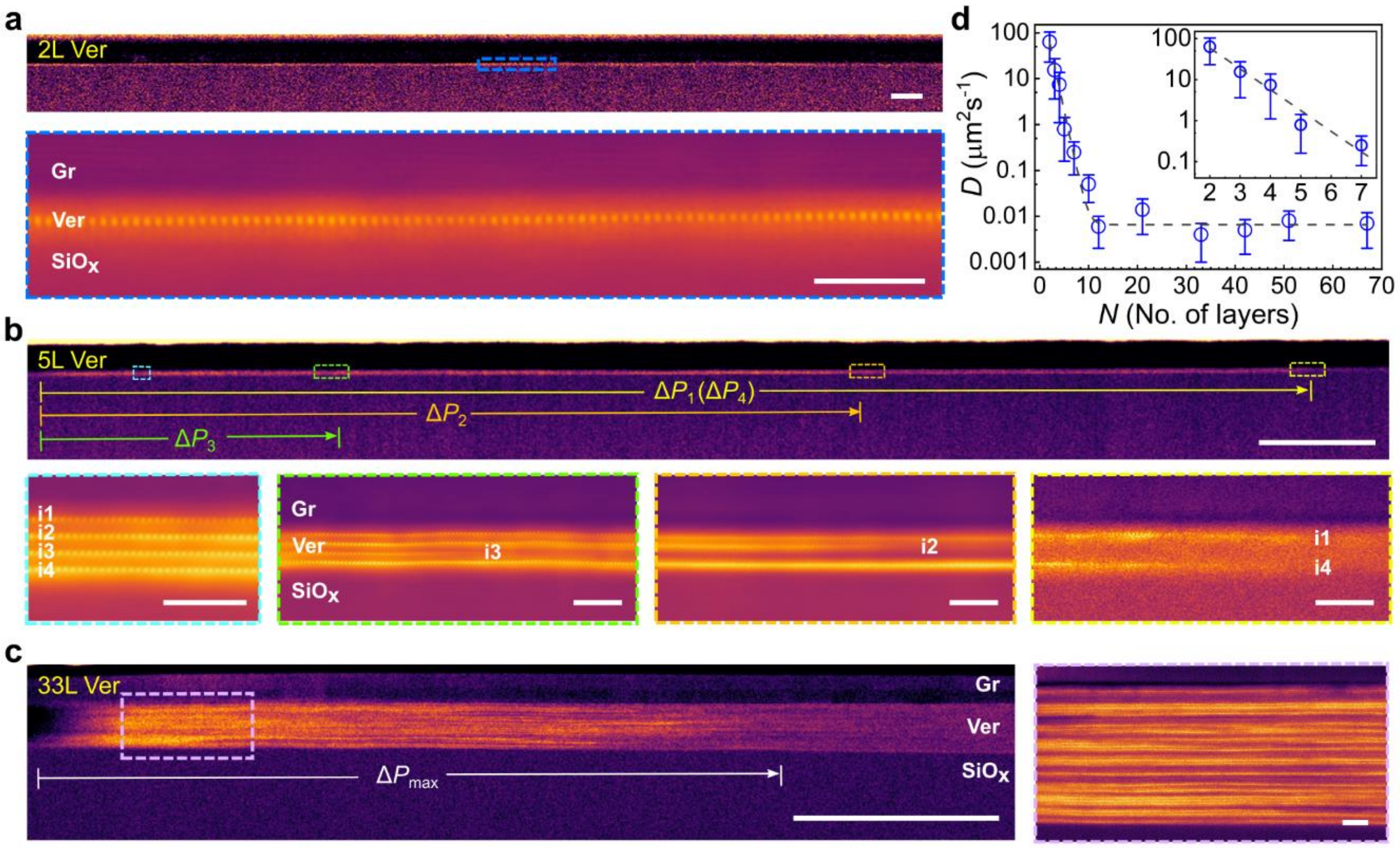

**Figure 2| Measuring the interlayer $Cs^+$ diffusivity for vermiculite flakes of different thickness**. **a**, Top panel, cross-section ADF-STEM image of 2L vermiculite (*t*=1 s). For this specimen (35 µm wide), $\Delta P \approx$ 14.5 µm (see Fig. S10 for optical image), which is too long to display in this STEM figure. Scale bar, 200 nm. Bottom panel, magnified view showing a bright line of Cs ion columns. Scale bar, 5 nm. **b**, Top panel, low magnification cross-section ADF-STEM image of a 5L vermiculite (*t*=10 s), with the left side showing the end of the flake where the $Cs^+$ penetration starts. Scale bar, 200 nm. Bottom panels, magnified images from the regions marked by dashed rectangles in the top panel. Scale bar, 5 nm. **c**, Left panel, cross-section ADF-STEM image taken from a 33L vermiculite exchanged for 10s. $\Delta P_{max}$ denotes the largest penetration distance among interlayers (for detailed penetration distances, see SI Tables S3.1-S3.15). Scale bar, 200 nm. Right panel, corresponding high magnification image. Scale bar, 5 nm. **d,** *D* as a function of *N*, measured from over 30 vermiculite flakes. Error bars, standard deviation from different samples. Inset, *D*(*N*) data for $N \leq 7$. Dashed lines, analytical formula described in the main text.

To understand these findings, we note that diffusion of interlayer ions in bulk clays depends strongly on interlayer expandability[16,22]. This suggests that this property could be responsible for the fast exchange observed in thin (<10 layer) clays. To investigate this hypothesis, we measured the swelling of our samples in liquid environment (0.1 M NaCl solution) using an atomic force microscope (Fig. S16). The interlayer expansion was found to be largest for bilayer flakes (~5 nm) and decayed rapidly with the number of layers, such that a stable value is reached for $N \geq 7$ layers (<0.5 nm expansion per interlayer). This strong thickness-dependent swelling behaviour mirrors that found in our diffusivity data and is consistent with previous swelling studies using other clays[22]. From a theory perspective, the interlayer expansion of clays has been proposed to depend on a balance of short-range Coulomb repulsion and long-range van der Waals attraction between the aluminosilicate layers[22,23]. The long-range interaction depends on the number of layers in the crystal, so this force decreases with decreasing *N*. This leads to a weaker attraction force between the layers, larger interlayer expandability and faster ion diffusivity in atomically thin samples. Our calculations of these van der Waals interactions for our samples are consistent with this interpretation (see SI Section 12.1).

**$Cs^+$ ion superlattices in twisted biotite mica**

An additional degree of freedom of atomically thin clays and micas is that these can restack in such a way that the crystal lattices of two neighbouring aluminosilicate planes are misaligned or 'twisted' with respect to each other (Fig. 3a,b). Unexpectedly, we find that within the interface of twisted samples, the exchanged ions arrange in an unusual pattern when observed at atomic resolution. Fig. 3 shows plan view characterization results from two Cs-exchanged biotite samples, each consisting of two monolayer crystals restacked with a small rotational misorientation. The twist angle between the upper and lower crystals, $\vartheta$, can be measured from the electron diffraction patterns (Fig. 3c,d) determined as $\vartheta \approx 9.5°$ and 2.6° for specimens shown in the top and bottom panels of Fig. 3, respectively. We find that the $Cs^+$ ions (bright dots shown in Fig. 3e-h) exchanged at the interface between these twisted bilayers arrange with the same in-plane interionic separation of 5.3 Å seen in the aligned bilayer samples. However, unlike in the arrangement seen in the aligned samples (Fig. 1b,d), these $Cs^+$ ions form groups of 7±2 and 130±15 atoms per island for $\vartheta \approx 9.5°$ and 2.6°, respectively (Fig. 3g,h). Outside these islands, $Cs^+$ is rarely observed. The islands occur periodically with periods of $L$ = 3.2±0.2 nm for $\vartheta \approx 9.5°$ and 11.7±0.7nm for $\vartheta \approx 2.6°$ (Fig. 3e,f). Interestingly, the found $L$ corresponds to the moiré periodicity of the underlying twisted aluminosilicate bilayers, $L_M$. This latter periodicity is given by $L_M = a/[2\sin(\vartheta/2)]$, where $a$ is the aluminosilicate lattice constant. For biotite, the known $a$ = 0.53 nm yields $L_M$ = 3.2 nm for $\vartheta \approx 9.5°$ and 11.7 nm for $\vartheta \approx 2.6°$ in excellent agreement with the measured periodicities, indicating that the $Cs^+$ ion islands are commensurate with the biotite moiré 'template'.

This behaviour is the result of the interaction of $Cs^+$ ions with the moiré superlattice. In general, ions bind into the local minima in the potential energy surface provided by the aluminosilicate layer. In a pristine bilayer mica (Fig. 1b), the top and bottom Al-Si-O hexagonal rings are aligned (AA stacking), resulting in a periodic array of energetically favourable hexagonal centre sites that host interlayer cations. In a twisted biotite sample, two potential energy surfaces superpose to form a new energy landscape with the larger period of the aluminosilicate moiré superlattice (Fig. S21). The moiré supercell contains various stacking sequences including high-symmetry AA and AB sites, as well as low-symmetry incommensurate bridging sites (Fig. 3a,b). The $Cs^+$ islands are aligned with the regions of AA stacking, equivalent to that present in perfect biotite and correlate with the favourable low energy regions in the potential energy landscape (Fig. S21, SI section 12.3). Notably, no $Cs^+$ is observed at AB regions, except for occasional $Cs^+$ bridges linking the islands *via* the incommensurate boundary regions (Fig. 3h). This suggests that these bridges are the route for $Cs^+$ propagation between the islands. Interestingly, the interlayer binding in the twisted interlayer appears to be weaker compared to the pristine interlayer (Fig. S21, SI Section 12.3). This is evident from the fast ion exchange in the twisted specimens, similar to what we observed for few-layered vermiculite samples. Pristine biotite crystals show no measurable ion exchange within a timescale of weeks (Fig. S12, SI Section 8), but twisted biotite crystals exhibit widespread interfacial ion exchange within 1 hour (Fig. S13). Here the thickness of the individual biotite crystals is not significant and we observed the same enhancement of Cs exchange at the interface between any two twisted biotite flakes – regardless of the precise twist angle and whether they were monolayers or much thicker ($\gtrsim$10 layer) flakes (see Fig. S15).

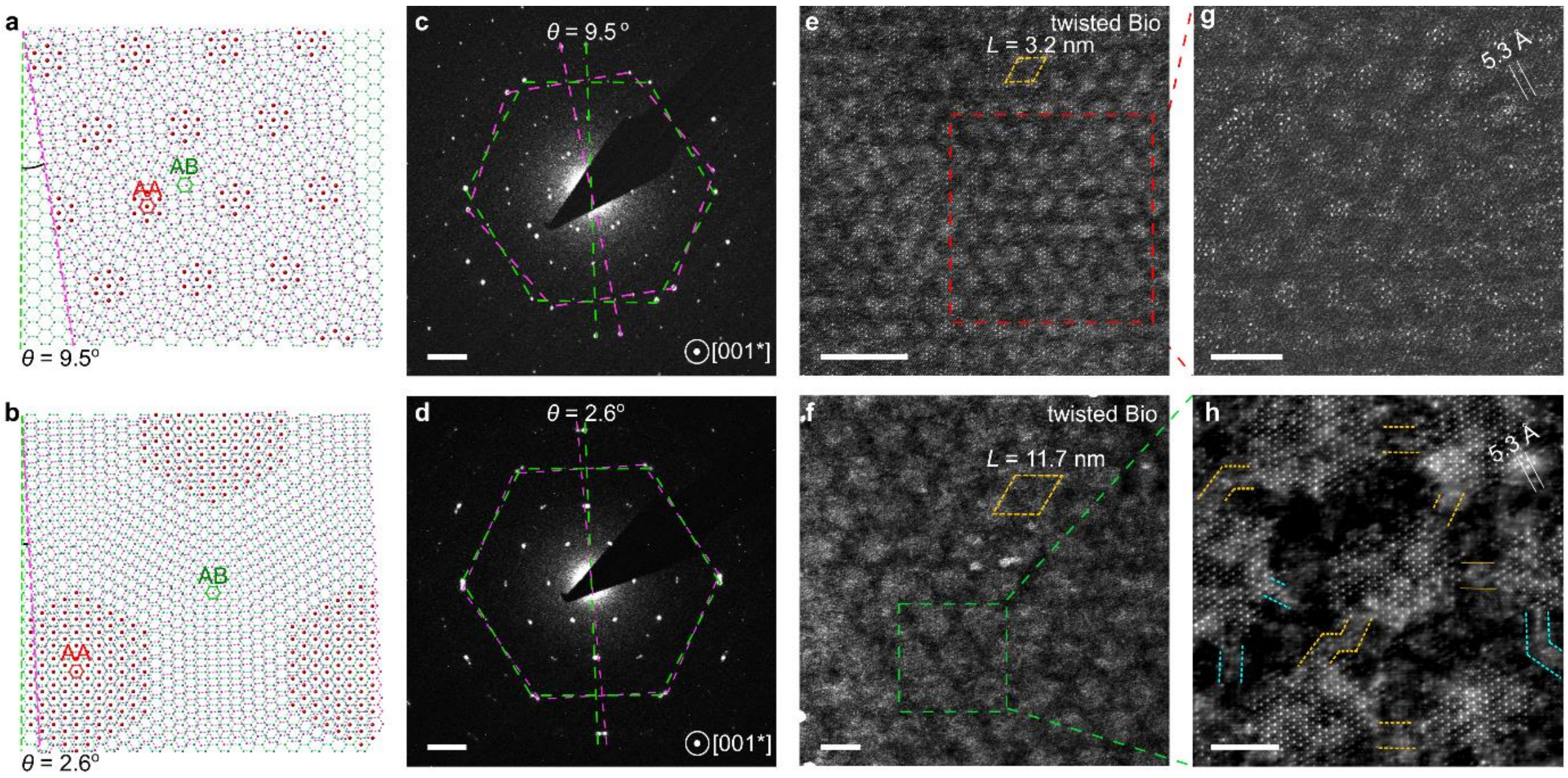

**Figure 3|Interlayer Cs islands in twisted biotite bilayers. a**,**b** Schematic atomic models of moiré patterns from twisted biotite monolayers for twist angles of $\vartheta \approx 9.5^{o}$ and $\vartheta \approx 2.6^{o}$ respectively. The two twisted aluminosilicate layers are shown in green and magenta respectively, with red balls showing the location of interlayer Cs (AA sites). The high-symmetry sites in this rigid lattice model are labelled as AA and AB according to the stacking sequence of aluminosilicate rings adjacent to interlayer ions. AA corresponds to the same stacking type as pristine bilayer biotite. **c,d,** Experimental selected area electron diffraction (SAED) patterns from twisted biotite bilayer samples, showing that the twist angles between the two biotite monolayers are $9.5^{o}$ and $2.6^{o}$ respectively. Scale bars, 2 $nm^{-1}$. **e,f,** Plan view ADF-STEM images taken from the Cs-exchanged twisted bilayer samples for $\vartheta \approx 9.5^{o}$ and $\vartheta \approx 2.6^{o}$, respectively. The $Cs^{+}$ ions form islands visible as brighter regions in the micrographs (see Fig. S14 for simulated image). The moiré unit cells are marked by the orange dashed obliques and the periodicities, *L,* are indicated. Scale bars, 10 nm. **g**,**h**, Magnified views from panels **e** and **f**, respectively. Interionic separation marked with white lines. Ion bridges between islands are marked by orange and blue dashed lines respectively, depending on their lattice orientation alignment. Scale bars, 5 nm.

**Ion exchange on the basal surface plane of muscovite mica**

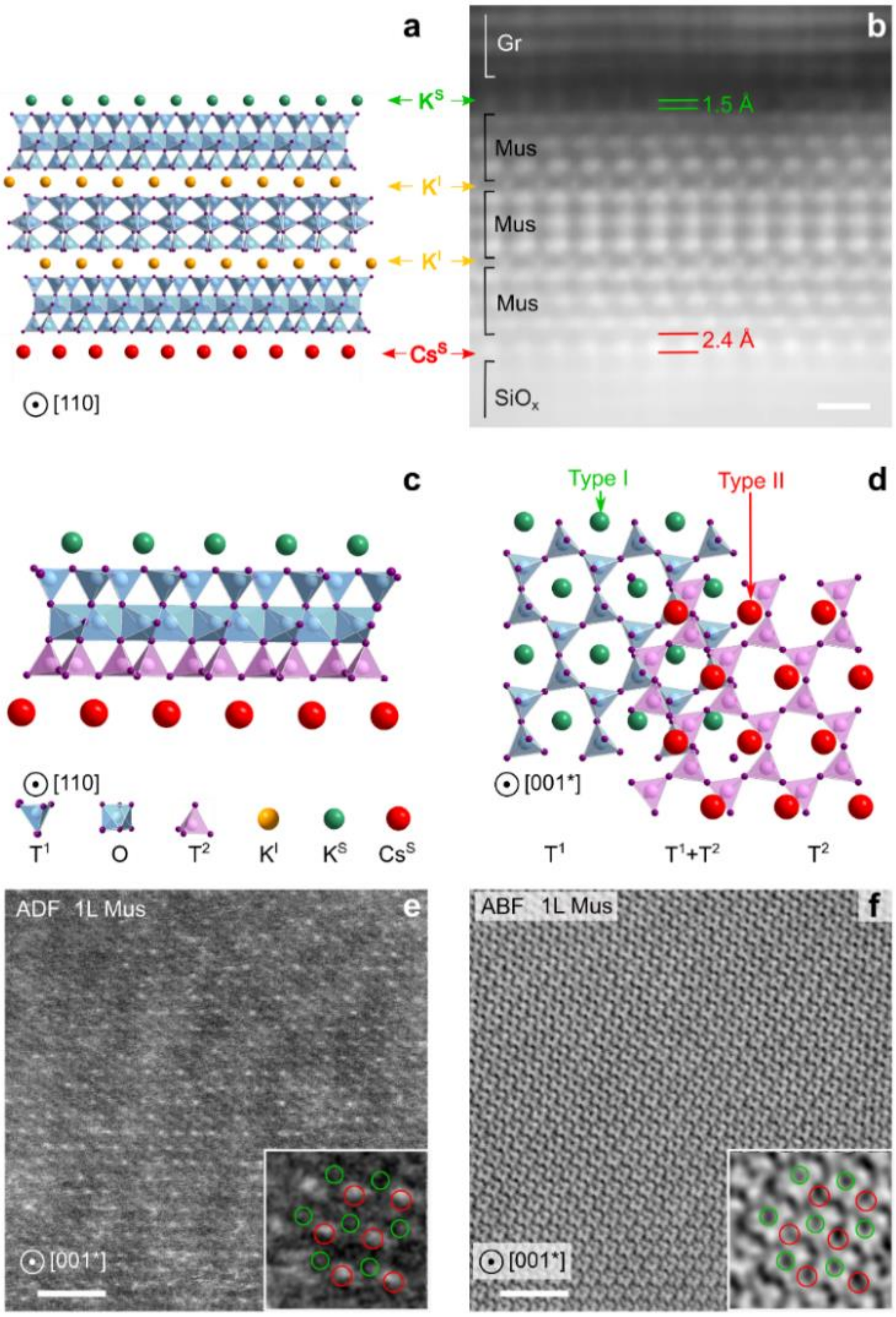

**Figure 4| Cations adsorbed on the surface basal plane of muscovite. a,b,** Schematic and corresponding cross-sectional ADF-STEM image of a 3L muscovite sample (see Fig. S8b for simulated image, Fig. S17b for raw image). The sample was encapsulated between graphite (Gr) and $SiO_x$. Distance between the surface ions and the outer aluminosilicate layer is marked with red horizontal lines. Superscripts 'S' and 'I' refer to surface and interlayer ions, respectively. Scale bar, 5 Å. **c**, Cross-sectional schematic of a monolayer muscovite with $K^+$ and $Cs^+$ ions adsorbed on opposite surfaces. The aluminosilicate TOT layers are indicated with the top ($T^1$) and bottom ($T^2$) tetrahedral sheets in the aluminosilicate layer coloured blue and magenta, respectively, to emphasize that the layers are not vertically aligned. **d**, Plan view schematic of the monolayer muscovite in panel **c**. For clarity, only the T sheets are shown (octahedral sheet is not shown). Type I and Type II refer to the different adsorption sites of $K^+$ and $Cs^+$ ions with respect to the neighbouring T-layer. For clarity, each of the two T sheets in the aluminosilicate monolayer is presented separately and then overlapped ($T^1+T^2$). **e**, ADF-STEM image of Cs-exchanged monolayer muscovite. Cs ions are visible as bright spots. **f**, Annular bright field (ABF)-STEM image of the same area imaged in panel **e**. Atomic positions are visible as dark spots. For both **e** and **f** higher magnification images are inset with red and green circles indicating the positions of $Cs^+$ and $K^+$ ions, respectively. Note that the $K^+$ ions are not identifiable in

ADF-STEM due to their low atomic number, so their position was obtained from the ABF image. Scale bars, 2 nm.

Characterising basal plane surface ions by STEM is experimentally more demanding and such studies have remained conspicuously absent in the literature. Nonetheless, our ultra-thin muscovite samples now allow for this study. We found that ADF-STEM of our cross-sectional samples can characterise the out-of-plane distance for native and $Cs^+$ exchanged ions, which was ~1 Å larger for $Cs^+$ than for the native ions (Fig. 4a,b, Fig. S17). Complementary plan view STEM imaging allows us to characterise with sub-angstrom resolution the in-plane ordering of surface $Cs^+$ ions and, crucially, the relative position of the ions with respect to the aluminosilicate backbone. To achieve this, we employ simultaneous ADF-STEM and annular bright-field (ABF)-STEM imaging[24]. The ADF image captures the heavy $Cs^+$ ions while the ABF image captures the lighter cations and the atoms in the aluminosilicate layer. Fig. 4e shows that surface $Cs^+$ ions, visible as bright spots in ADF, form a quasi-hexagonal lattice. The ABF image of the same area is shown in Fig. 4f. Here, atomic positions are clearly visible as dark spots, which indicates that the specimen remained crystalline during imaging (see also Fig. S18, Fig. S19). This image shows the expected Al-O and Si-O ring structures, known in the literature as hexagonal or ditrigonal rings[12] (coloured blue and magenta in Fig. 4d). Correlation of the ADF and ABF-STEM images reveals that $K^+$ and $Cs^+$ ions are adsorbed on different sites within these rings and, in this sample, on opposite surfaces. $K^+$ ions adsorb in the centre of the hexagonal rings (Type I site) on the top surface ($T^1$); whereas statistically $Cs^+$ ions are found to overwhelmingly (76%, see Fig. S18) adsorb on the vertex of the hexagonal rings (Type II site on $T^2$). The difference between in-plane and out-of-plane coordinates observed here for $Cs^+$ compared to the native $K^+$ cations is attributed to the presence of multiple hydration complexes for $Cs^+$ ions[6, 25-27]. These lead to different favourable adsorption sites during the exchange process in solution that are preserved when imaging (see Fig. S20 and SI Section 12.2).

## Conclusions

Our results provide atomic-scale insights into surface ion adsorption and interlayer ion diffusion in micas and clays. This is of relevance for optimising novel clay-membrane designs[1-6] and for understanding the limited movement of heavy metal ions in contaminated land[17]. Beyond this, our STEM 'snapshots' method provides fundamentally new insights into ion transport in highly confined spaces, of interest following recent developments[28]. Given the wide research interest in the electronic and optical properties of 2D metals[29] and in twisted 2D materials heterostructures[30], our observations suggest opportunities for fabricating mica-encapsulated 2D metal ion superlattices[31]. Finally, our demonstration of enhanced ion exchange speed and capacity in few-layer crystals provides evidence that exfoliated clays and micas can be used to produce membranes, filters and adsorption products with enhanced performance.

## Acknowledgments

The work was supported by EPSRC grants EP/M010619/1, EP/P00119X/1, EP/S021531/1 and EP/P009050/1, the European Research Council (ERC) under the European Union's Horizon 2020 research and innovation programme (Grant ERC-2016-STG-EvoluTEM-715502 and the ERC Synergy Hetero2D project 319277) and The Royal Society. L.M. acknowledges the EPSRC NOWNano programme for funding. Part of this work was supported by the Flemish Science Foundation (FWO-Vl).

The authors declare no competing interests

All raw data is available from the corresponding authors on reasonable request

## Ion exchange in atomically thin clays and micas

**Supplementary Information**

Yi-Chao Zou[1,2], Lucas Mogg[3,4,5], Nick Clark[2,3], Cihan Bacaksiz[6], Slavisa Milanovic[6], Vishnu Sreepal[3,7], Guang-Ping Hao[3,4], Yi-Chi Wang[2,8], David G. Hopkinson[2,3], Roman Gorbachev[3,4], Samuel Shaw[9], Kostya S. Novoselov[3,4], Rahul Raveendran-Nair[3,7], Francois M. Peeters[6], Marcelo Lozada-Hidalgo[3,4]*, Sarah J. Haigh[2,3]*

[1]School of Materials Science and Engineering, Sun Yat-sen University, Guangzhou, 510275, P. R. China
[2]Department of Materials, The University of Manchester, Manchester M13 9PL, UK
[3]National Graphene Institute, The University of Manchester, Manchester M13 9PL, UK
[4]Department of Physics and Astronomy, The University of Manchester, Manchester M13 9PL, UK
[5]Department of Engineering, University of Cambridge, 9 JJ Thomson Avenue, Cambridge CB3 0FA, UK
[6]Departement Fysica, Universiteit Antwerpen, Groenenborgerlaan 171, B-2020 Antwerp, Belgium
[7]Department of Chemical Engineering and Analytical Science, The University of Manchester, Manchester, M13 9PL, UK
[8]Beijing Institute of Nanoenergy and Nanosystems, Chinese Academy of Sciences, Beijing 101400, P. R. China
[9]Research Centre for Radwaste Disposal and Williamson Research Centre, School of Earth and Environmental Science, The University of Manchester, Manchester M13 9PL, UK

# Content

## 1. Source Materials

The materials studied in this work were: ruby muscovite, purchased from Agar Scientific; natural biotite, sourced from Lanark, Ontario (Canada); and a natural vermiculite clay sourced from Gorainas, Minas Gerais (Brazil). The chemical composition of the three materials was provided by the suppliers (Table S1). The simplified crystal structure, approximate unit cell dimensions and chemical formula for the materials are given in Table S2.

**Table S1|** Material composition (wt. %), as received from supplier list.

| Material | $SiO_2$ | $Al_2O_3$ | $K_2O$ | MgO | $Fe_2O_3$ | Water |
|---|---|---|---|---|---|---|
| Muscovite | 45.57 | 33.10 | 9.87 | 0.38 | 2.48 | 2.99 |
| Biotite | 50.79 | 18.45 | 3.72 | 12.28 | 3.21 | 4.9 |
| Vermiculite | 40.4 | 11.1 | 0.01 | 26.9 | 7.85 | 11.43 |

**Table S2|** Indicative crystal structure and unit cell dimensions of the source materials[1-3]. Chemical formulas are expressed in the form of **interlayer cation, octahedral layer and tetrahedral layer with associated -OH, and water molecules.**

| Crystal | a(Å) | b(Å) | c (Å) | $\alpha$ (°) | $\beta$(°) | $\gamma$(°) | Space group | Chemical formula |
|---|---|---|---|---|---|---|---|---|
| Muscovite | 5.2 | 9.0 | 20.0 | 90.0 | 95.2 | 90.0 | c1(15) | $KAl_2AlSi_3O_{10}(OH)_2$ |
| Biotite | 5.3 | 9.2 | 20.1 | 90.0 | 95.1 | 90.0 | c1(15) | $K(Mg, Fe)_3AlSi_3O_{10}(OH)_2$ |
| Vermiculite | 5.3 | 9.1 | 28.9 | 90.0 | 97.1 | 90.0 | c1(15) | $MgMg_3(Si, Al)_4O_{10}(OH)_2 \cdot 4H_2O$ |

## 2. Transmission electron microscopy (TEM) and specimen preparation

### 2.1 Specimen preparation

TEM sample preparation is undertaken (in a class 100 clean room environment) as follows. First, fresh mica or vermiculite surfaces are exposed from bulk crystals via cleavage using clean scalpel blades. This minimises accumulated contaminants on the topmost surfaces. Next, this freshly exposed surface is further cleaved with adhesive back-grinding tapes (Nitto BT-50E-FR, Nitto Denko tape selected for high adhesive retention and high cleanliness). The crystals were peeled several times onto the tape until a good coverage can be seen. The tape covered with the flakes is then swiftly pressed against either an $O_2$/Ar plasma cleaned $SiO_x$ substrate (for cross-sectional samples), or polyvinyl alcohol (PVA) - polypropylene carbonate (PPC) coated Si wafer (for plan view samples). Next, candidate flakes of desired thicknesses are optically identified (Fig. S1a). Thereafter, the thickness and cleanliness of the target flakes are confirmed via atomic force microscopy (AFM) (Fig. S1b). After this stage, the fabrication process diverges for cross-sectional and plan view sample preparation, as described below.

The focussed ion beam milling (FIB) method is used to prepare cross-sectional TEM samples (Fig. S2a-d). The $SiO_x$/Si substrate with the target flakes is immersed in 0.1 M $CsNO_3$ solution for the desired amount of time for ion exchange. The flakes are then cleaned in DI water and dried in a nitrogen atmosphere, where the flake surface is found to stay optically clean after the ion exchange process (Fig. S2c,d, Fig. S10). To avoid surface damage during FIB sample preparation, the target clay (or mica) flake is protected by covering it with a graphite flake. This process is based on a dry transfer method[4] using PVA-polymethyl methacrylate (PMMA) as the carrier layer. Note that only the top surface of the graphite had contact with the transfer polymer, which keeps the interface (bottom surface of graphite) of the graphite and the target clay (or mica) flake clean. Next, the $SiO_x$-flake-graphite assembly is cleaned using the following method: acetone (120 s) → isopropyl alcohol (120 s) → DI water (120 s) → $N_2$ dry (60 s). This sequence is repeated until the sample is optically free from contamination (typically, one cycle of cleaning is enough). The $SiO_x$-flake-graphite assembly is then protected by depositing a narrow Pt strip (2 µm thick, 1 µm wide and 8-40 µm long) inside the FIB scanning electron microscope (FIB SEM, FEI Helios 660). The Pt strip is deposited perpendicular to crystallographic edges of the flakes observed under an optical microscope and SEM, to assist with orienting the crystal during TEM imaging. Standard milling protocols[5] are used to cut the lamella free from the substrate and transfer it to a pillar of a specialist OmniProbe™ Cu TEM support grid, followed by 30 kV, 16 kV, 5 kV and 2 kV ion beam milling and polishing to electron transparency (see Fig. S2a-d).

For plan-view specimens, target mica or clay flakes exfoliated on PVA/PPC are transferred onto $SiN_x$ TEM grids, using the dry transfer method[4] followed by the cleaning cycle described above (acetone (120 s) → isopropyl alcohol (120 s) → DI water (120 s) → $N_2$ dry (60 s). The ion exchange is then conducted by immersing the whole TEM grid (containing the target flake) into a 0.1 M $CsNO_3$ solution for the desired time length (Fig. S2g), following by cleaning using ethanol and DI water. It is then dried and stored in $N_2$ atmosphere prior to TEM analysis (see Fig. S2h). For the fabrication of twisted mica flake samples, we identify twisted regions in the as-exfoliated flakes on $SiO_x$ substrates using optical microscopy. The target flake with twisted regions is then transferred onto the $SiN_x$ TEM grids as described above, followed by ion exchange in 0.1M $CsNO_3$ water solution for the desired time length. Both cross-section and plan-view specimens are stored in $N_2$ atmosphere prior to TEM measurements.

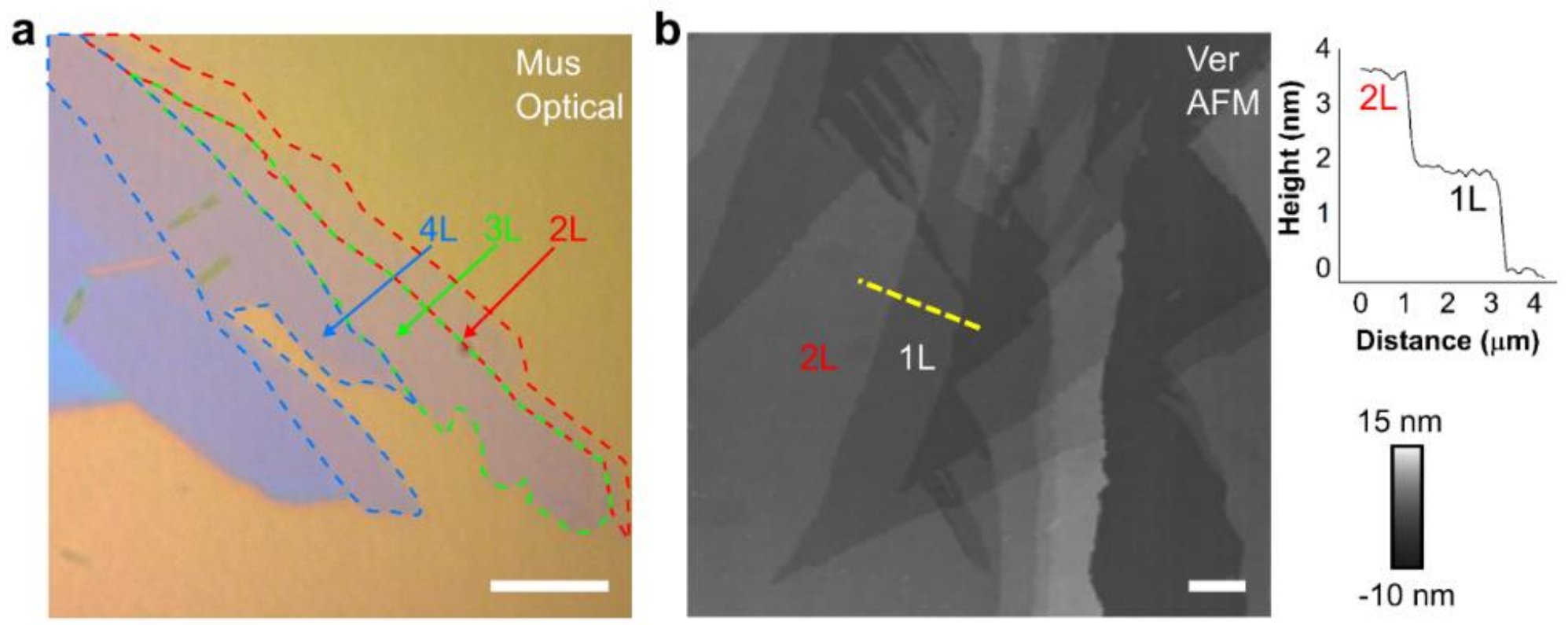

**Supplementary Figure S1 | Identifying atomically thin crystals. a,** Optical image of a mechanically exfoliated muscovite flake on a $SiO_x$/Si substrate. Scale bar, 10 μm. **b,** AFM image of a typical exfoliated vermiculite flake in ambient air. Right-top inset, the height profile of the section marked by the yellow dashed line marked in **b**, confirming that the height of the monolayer (bilayer) region is ~1.8 nm (~3.5 nm). Scale bar, 1 μm.

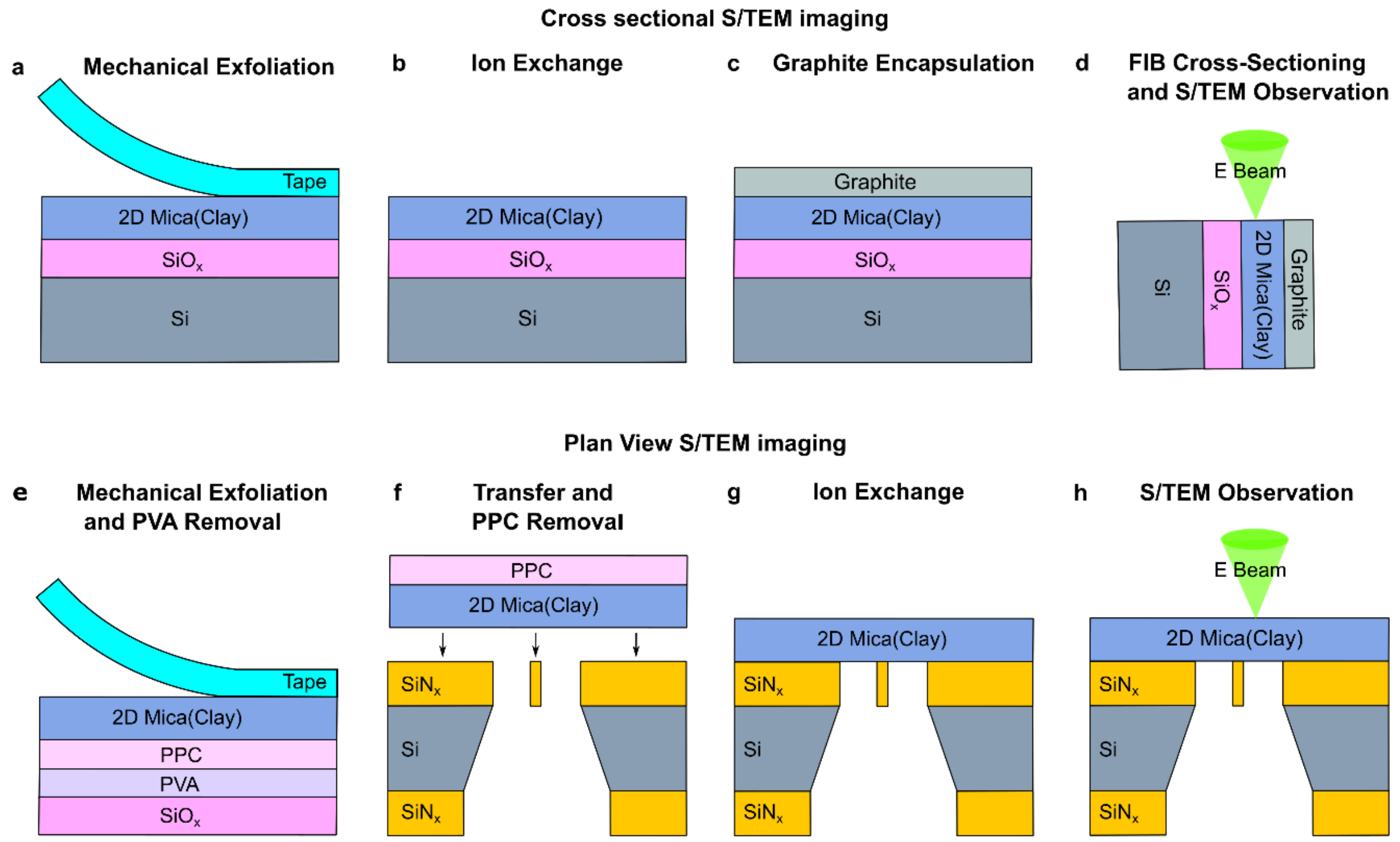

**Supplementary Figure S2 | Schematic illustrating the procedure of TEM sample preparation. a-d**, Cross-sectional samples. **a**, Mechanically exfoliated mica(clay) flakes onto $SiO_x$ (90-290 nm)/Si substrate. **b**, Ion exchange processes for the selected flakes: electrolyte immersion; then drying in a $N_2$ gas environment and quality inspection by optical microscopy. **c**, Cover the target flakes with thin graphite to avoid surface damage or contamination. **d**, FIB lift out of the graphite/$SiO_x$/mica(clay) stack, followed by FIB thinning and S/TEM imaging. **e-h**, Plan-view samples. **e**, Mechanical exfoliation of mica(clay) crystals onto a PVA/PPC coated $SiO_x$ (90-290nm)/Si substrate, followed by dissolving PVA in water to leave a floating PPC/mica(clay) stack. **f**, Transfer of the PPC/mica(clay) stack by micromanipulation onto a $SiN_x$/Si TEM grid, followed by removal of PPC in solvent to leave a clean and suspended flake. **g**, Ion exchange and cleaning of the mica(clay) flake on TEM grid. **h**, STEM and TEM imaging.

### 2.2 Scanning Transmission Electron Microscope (STEM) and TEM imaging

A probe-side aberration-corrected FEI Titan G2 80-200 S/TEM “ChemiSTEM” microscope was used for annular dark field (ADF) scanning transmission electron microscopy (STEM). This microscope was operated at 200kV with a probe current of ~6-15 pA, a convergence angle of 21 mrad and a ADF detector with an inner (outer) collection angle of 48(196) mrad. For STEM energy dispersive X-ray spectroscopy (EDS), we used a Super-X EDS detector with a beam current of ~6 pA, a per pixel dwell time of 50 µs, a total acquisition time of 100 s and a pixel size of 0.05 nm. Selected area electron diffraction (SAED) patterns were obtained using the same microscope operated in TEM mode. For the FIB lamellae, SAED patterns were used to orientate the specimen along low-index zone axis using low electron fluence in order to achieve atomically resolved high-resolution STEM image data.

A JEOL ARM300CF double aberration-corrected microscope was used to acquire ADF- and annular bright field (ABF)-STEM images simultaneously. This instrument has a cold FEG electron source, was operated at 80 kV with a probe current of ~6 pA and a convergence angle of 24.8 mrad. Images were acquired with a ADF detector having a collection angle of ~50-150 mrad and an ABF detector having a ~12-24 mrad collection angle. Multi-slice image simulation for high-resolution STEM images was conducted using atomic models created under the Atomic Simulation Environment[6], QSTEM software[7] with the above experimental parameters, and a source size of 0.8-1 Å.

### 3. XRD, TEM and STEM Characterization of pristine crystals

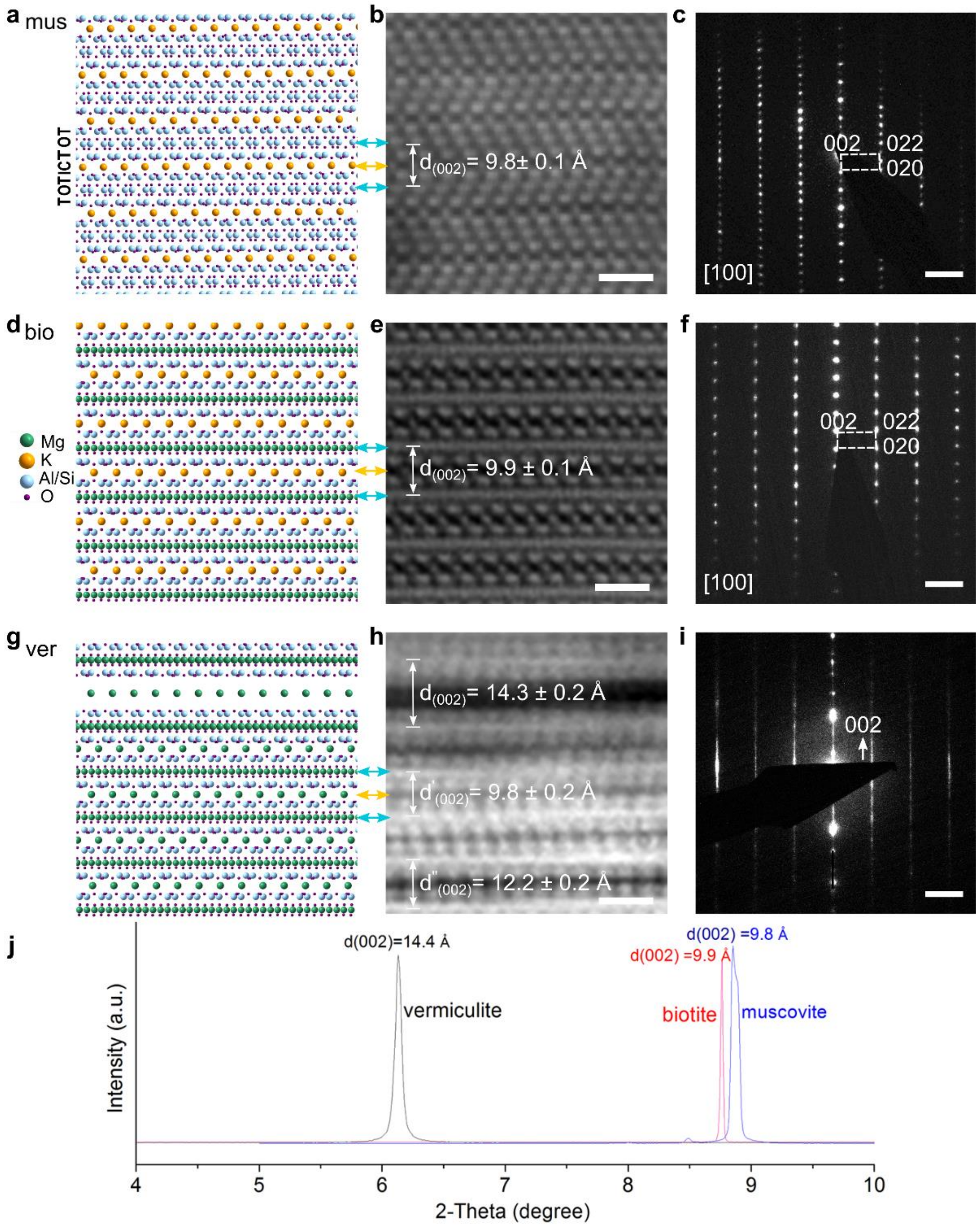

**Supplementary Figure S3|S/TEM and XRD characterization of bulk crystals.** Atomic model, cross-section STEM micrograph and diffraction patterns for **a-c,** muscovite; **d-f,** biotite; and **g-i,** vermiculite. **a,d,g,** Schematic atomic models showing the crystal structures of the source materials. **b,e,h,** Cross-sectional annular dark field (ADF) STEM images. Scale bars, 1 nm. All materials consist of alternating layers of aluminosilicate (TOT) and interlayer cations (IC) when viewed along the [100] direction. The atomic models have been aligned to the experimental images with some octahedral 'O' layers and 'IC' layers highlighted by blue and orange arrows, respectively. **c,f,i,** Selected area electron diffraction (SAED) patterns, for the materials viewed along [100]. Scale bars, 2 $nm^{-1}$. **j,** Powder X-ray diffraction (XRD) spectra taken from the source crystals, showing that each pristine crystal has uniform interlayer spacing under ambient environmental conditions.

The as received bulk crystals were characterized in house using cross-sectional STEM imaging, SAED and X-ray diffraction (XRD) (Fig. S3). XRD measurements (Fig. S3j) reveal the expected differences in (002) interlayer separations for the three source crystals. Pristine bulk specimens were found to have interlayer spacings of 9.8, 9.9 and 14.4 Å for muscovite, biotite and vermiculite respectively, when measured under ambient environmental conditions; consistent with expected literature values[1-3]. **The differences in these interlayer spacings can be related to the ability of the crystals to hold interlayer water; a property closely related to their composition and interlayer ion species (Section 1, Table S1).** The exchangeable ions in all three materials are typically $K^+$ or $Mg^{2+}$. From Table S1, S2, it can be seen that the interlayer cation is dominated by $Mg^{2+}$ in vermiculite, by $K^+$ in muscovite, and by $K^+$ in biotite. Compared to $K^+$, $Mg^{2+}$ has a much lower hydration enthalpy[8], and hence vermiculite has a higher capacity to hold interlayer water than muscovite or biotite and a correspondingly larger interlayer spacing. Indeed, the difference in interlayer ion species is visible from our ADF-STEM images where intensity is strongly dependent on atomic number. Hence, the native interlayer $K^+$ cations are directly observed for muscovite and biotite, but the lighter $Mg^{2+}$ native interlayer ions in vermiculite are below the signal to noise ratio of the image (see Fig. S3b, e, h, where the positions of the interlayer cations are indicated by yellow arrows).

After cross-sectional sample preparation and transfer to the S/TEM ultra-high vacuum environment, we measured uniform local interlayer spacing of 9.8 to 9.9 Å for non-exchanged muscovite and biotite (Fig. S3b,e), consistent with our XRD values (Fig. S3j). In contrast, for non-exchanged vermiculite we observe interlayer distances in the S/TEM ranging from ~9.8 Å to ~14.3 Å (Fig. S3h). This is attributed to the known variable water content in this crystals' interlayer space. Vermiculite is known to contain 0-3 molecular layers of water associated with its $Mg^{2+}$ interlayer cations, with the precise number of water molecular layers depending on the humidity of the surrounding environment[9,10]. This yields a variable interlayer spacing, which we observe in vermiculite samples under the TEM vacuum (Fig. S3h, Fig. S4). In Fig. S3h, the top interlayer space in the figure is 14.3 Å wide, which is consistent with the presence of two molecular layers of water. The same can be observed in the bilayer sample in Fig. S4. The bottom interlayer spacing in Fig. S3h is 12.2 Å wide, which corresponds to partial dehydration with one molecular layer of water; whereas the spacing of 9.8 Å indicates a fully dehydrated layer.

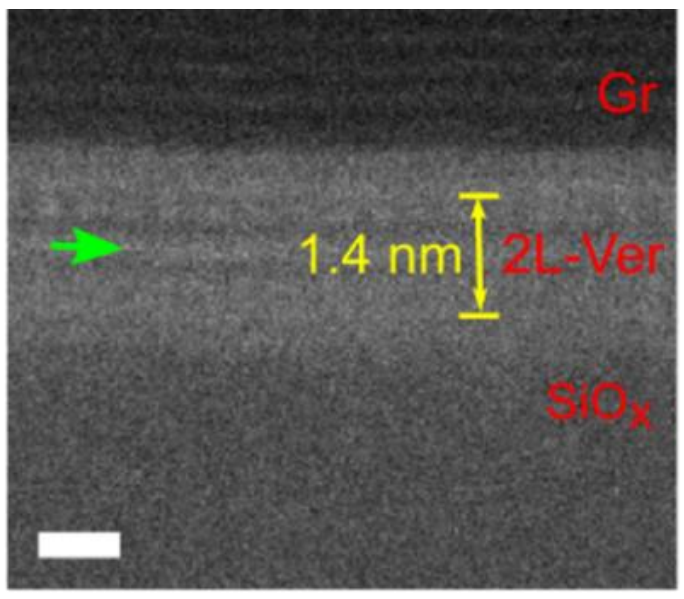

**Supplementary Figure S4|Characterisation of native interlayer spacing in cross-sectional sample of 2L vermiculite**. The green arrow marks the hydrated interlayer $Mg^{2+}$. Scale bar, 1 nm.

In addition to the differences in interlayer separation, ADF-STEM imaging also reveals differences in the octahedral (O) layers for the three materials. Substitution of $Al^{3+}$ in the O layer by other cations with lower valence charge (e.g. $Mg^{2+}$) changes the local co-ordination environment from a dioctahedral to a trioctahedral structure[11]. In our samples, the O layer of muscovite is dioctahedral, whereas vermiculite and biotite are trioctahedral. This difference is visible in the cross-sectional ADF-STEM images (Fig. S3b,e,h), where the sparser atomic arrangement in dioctahedral muscovite produces a lower ADF intensity within the O layer compared with that observed in trioctahedral vermiculite and biotite.

## 4. Identifying the interlayer $Cs^+$ ions by cross-sectional imaging

### 4.1 ADF-STEM Imaging

The atomic-number sensitivity of the ADF-STEM imaging mode can also be used to gain quantitative insights into occupancies for individual atomic columns. However, compared to what is achievable for model oxides, the interpretation of ion occupancy for clays (micas) has a relatively large error bar due to the complexity of the structure, variable hydration and site occupancies, as well as poorly defined Debye Waller factors. Moreover, the lower electron dose necessary to prevent structural damage in clays (micas), limits the achievable signal to noise ratio. Despite these limitations, it is informative to explore the potential of ADF-STEM for semi-quantitative insights into the interlayer ion exchange in these materials.

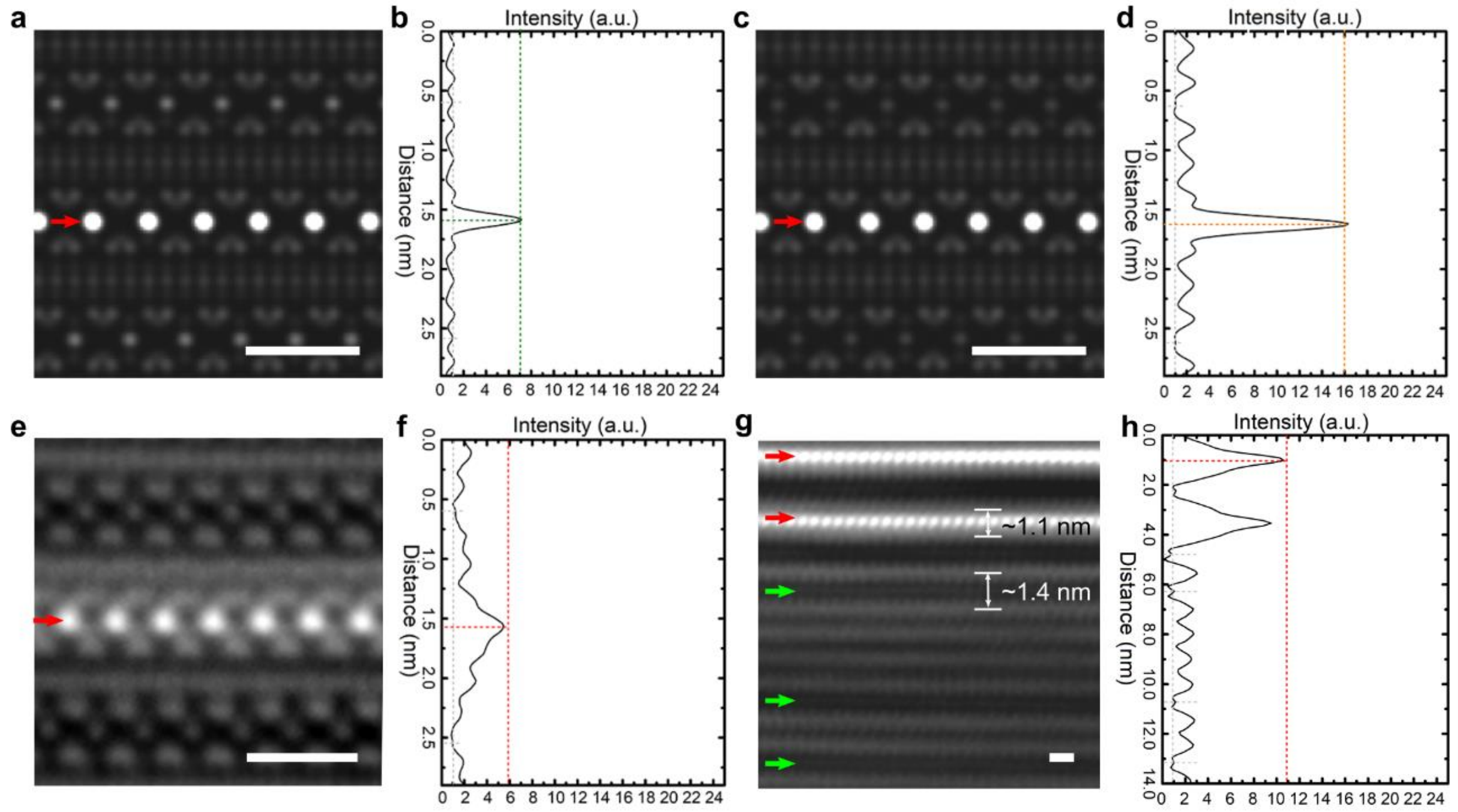

**Supplementary Figure S5| Interpreting Cs occupancy in the interlayer cation sites from ADF images. a,c,** Simulated ADF-STEM images for K-biotite (**a**) and dehydrated Mg-vermiculite (**c**) respectively. Image simulation used thermal diffuse scattering (TDS) of 25 iterations, with a supercell thickness of 2.1 nm, using simulation parameters given in Section 2.2. In the simulated structures, each only has one atomic plane that is fully exchanged for $Cs^+$, highlighted by the red arrows. **b,d,** Corresponding line intensity profiles summed horizontally across the whole image, normalized to the native interlayer cation column intensity. The intensity ratio between the exchanged layer and the native interlayer cation is found to be $I_{Cs}/I_K \approx 7$ for biotite and $I_{Cs}/I_{Mg} \approx 15$ for vermiculite. **e,g,** Experimental ADF-STEM images for Cs exchanged biotite (**e**) and vermiculite (**g**) respectively. Red arrows denote the Cs-exchanged atomic layer. Green arrows denote the hydrated interlayer $Mg^+$ layer. Please note that $Cs^+$ ion exchange in biotite is rare and confined to ~10 nm from the edges (see Fig. S12b). **f,h,** Corresponding line intensity summed horizontally across the whole image and normalized to the native ion column intensity. These show $I_{Cs}/I_K \approx 6$ for biotite and $I_{Cs}/I_{Mg} \approx 11$ for vermiculite. All scale bars, 1 nm.

For full ion exchange, simulations show that the ADF intensity of ion columns in an exchanged interlayer should have increased by a factor of 7 for $Cs^+$ substituting $K^+$ (as in K-biotite) and by a factor of 15 for $Cs^+$ substituting $Mg^{2+}$(as in Mg-vermiculite) as demonstrated in Fig. S5a-d. Partial ion exchange will result in a correspondingly smaller increase in ADF intensity. Fig. S5e shows an experimental ADF-STEM image taken from a Cs exchanged biotite. Ion exchange in biotite is rare and typically confined to a few nanometres from the flake edge. Nonetheless, biotite was chosen as its lower hydration means it is more robust to high-resolution imaging. From the image and line intensity profile (Fig. 5f), the intensity of the exchanged layer relative to an apparently pristine interlayer region, $I_{Cs}/I_K$, is measured to be around 6, very close to the theoretical value $I_{Cs}/I_K \approx 7$. This semi-quantitative analysis thus suggests this interlayer is fully exchanged. Lower $I_{Cs}/I_K$ values were measured in other data suggesting that partial exchange is also possible.

For vermiculite with $Mg^{2+}$ interlayer cations, experimentally determining the interlayer occupancy is much more challenging due to the relatively weak ADF-STEM signals of light $Mg^{2+}$, and the necessity for low dose imaging of these beam sensitive hydrated minerals (total electron fluence allowance ~$10^4$ e·$A^{-2}$ for Mg-vermiculite and ~$10^5$ e·$A^{-2}$ for dehydrated Cs-vermiculite) giving relatively large experimental errors[12]. Our qualitative analysis revealed a highest $I_{Cs}/I_{Mg} \approx 11$ from dozens of samples examined, as illustrated in Fig. S5g-h. This is smaller than the theoretical $I_{Cs}/I_{Mg} \approx 16$ and may suggest only partial occupancy of $Cs^+$ was observed. Nonetheless, we find that due to the material's electron beam sensitivity and complex chemistry, analysis of interlayer ion occupancies requires complementary plan view imaging as discussed in the main text.

### 4.2 BF STEM Imaging

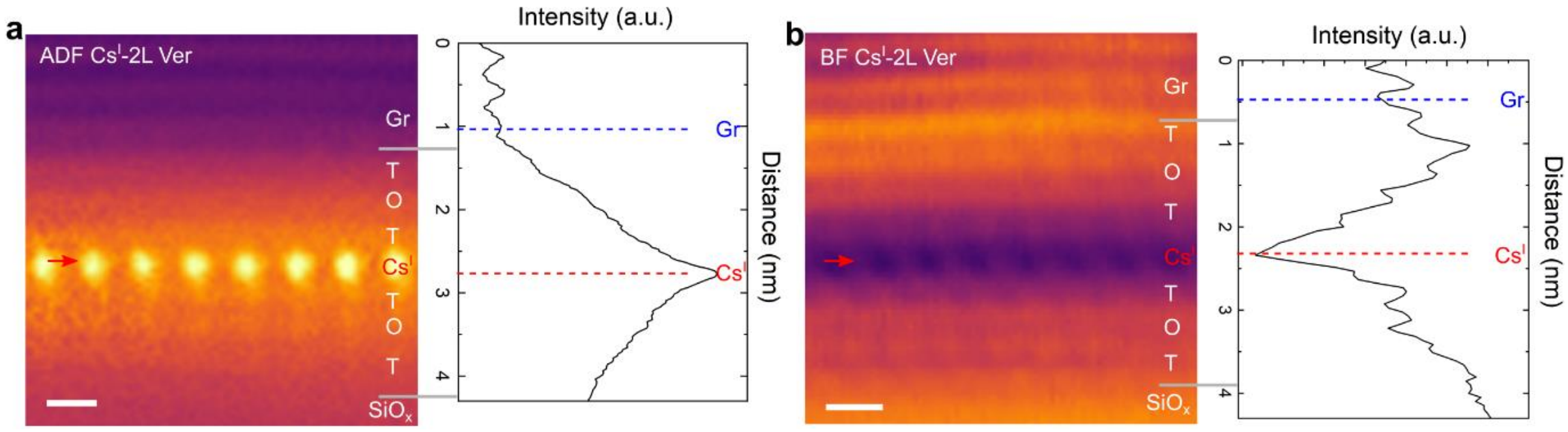

**Supplementary Figure S6| Combining ADF with BF STEM imaging to confirm that Cs occupies the interlayer spaces in a 2L vermiculite. a,** Left panel, ADF-STEM image. Right panel, corresponding line intensity profiles showing that the $Cs^+$ peak is located at the centre of the clay region between the two TOT layers. $Cs^I$-Gr distance of ~ 18 Å. **b,** Left panel, BF-STEM image, where the darkest dots indicating the Cs occupying sites. Right panel, corresponding line intensity profiles. Phase contrast in BF image can clearly show the TOT layered structure, confirming that Cs ions reside at the interlayer spaces between TOT layers. Scale bars, 5 Å.

## 4.3 STEM EDS

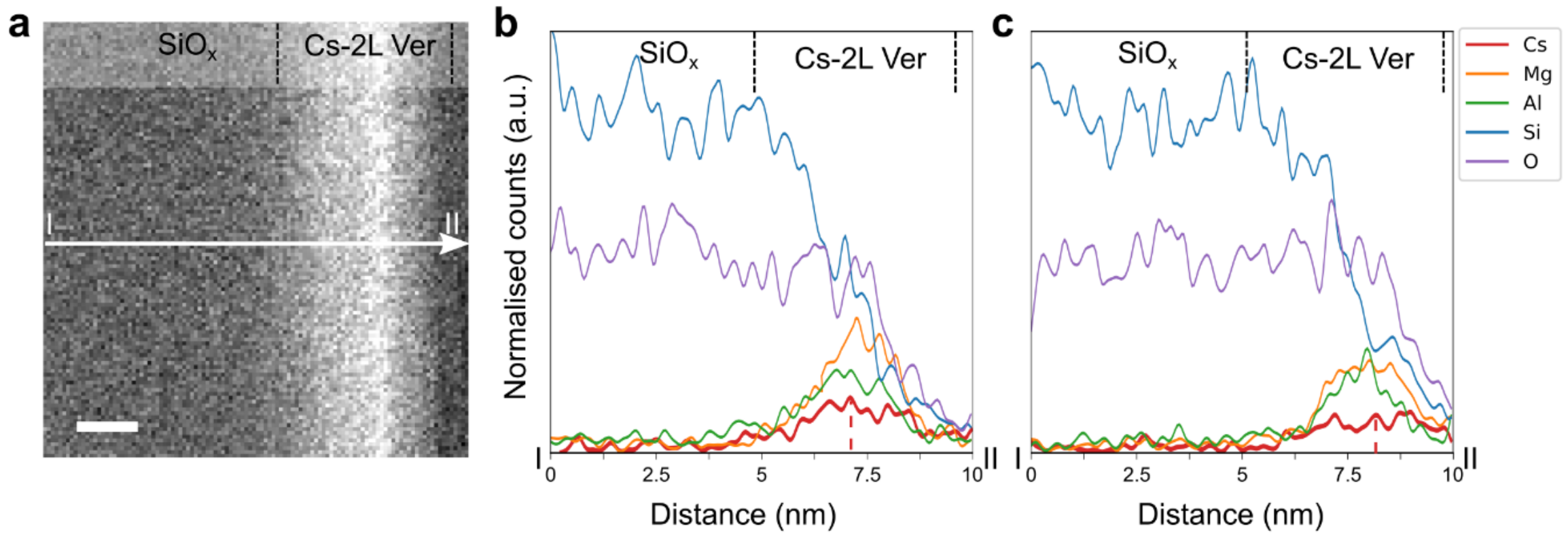

**Supplementary Figure S7| EDS characterization for beam sensitive clay/mica samples. a,** ADF-STEM cross-sectional image taken from a Cs exchanged 2L vermiculite sample. Scale bar, 1 nm. **b,c** Consecutively acquired EDS elemental line profiles along the line I to II (summed vertically for the whole ADF region shown in **a**). **b**, EDS line profile from $t$=0 to $t$=100 s with $t$ being the EDS acquisition time, total electron fluence of ~1.6x10$^7$ e·Å$^{-2}$. **c,** EDS line profile collected from $t$=100 to $t$=200 s with material having experienced a total electron fluence of ~3.2x10$^7$ e·Å$^{-2}$. The peak values for $Cs^+$ counts are marked by red dashed lines showing that $Cs^+$ diffuse away from the region of interest causing a lower peak Cs concentration after increased electron beam irradiation (Cs counts in **b** > **c**). A similar decrease is seen in the Mg concentration. Note that the total dose required for high-resolution STEM EDS elemental mapping is >10$^7$ e·Å$^{-2}$, much larger than the total electron fluence that would amorphize the Cs-vermiculite crystals (10$^5$ e·Å$^{-2}$), and hence it is not possible to gain atomic-scale chemical mapping or quantitative atomic analysis in these clay samples.

# 5. STEM Image interpretation for crystal structure

## 5.1 Crystal structure of muscovite

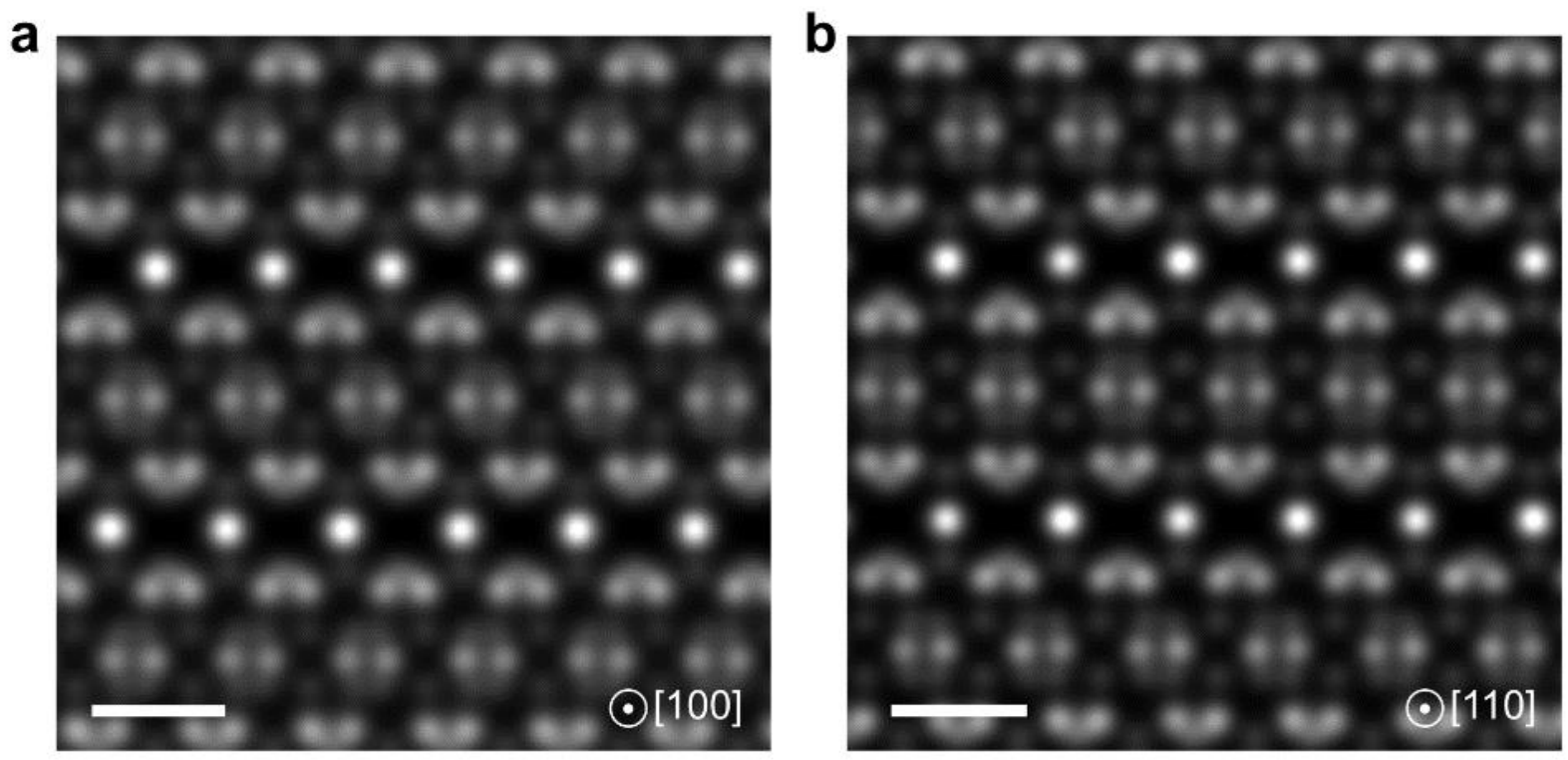

**Supplementary Figure S8|Simulated ADF-STEM cross-sectional images of muscovite. a,** Viewed along [100] zone axis. In agreement with experimental images in Fig. S3b, Fig. S17a. **b,** Viewed along [110] zone axis. In agreement with experimental images in Fig. 4b. Image simulation used TDS with 25 iterations, a supercell thickness of 8.0 nm, using the parameters given in Section 2.2. Scale bars, 5 Å.

## 5.2 Stacking of interlayer cations

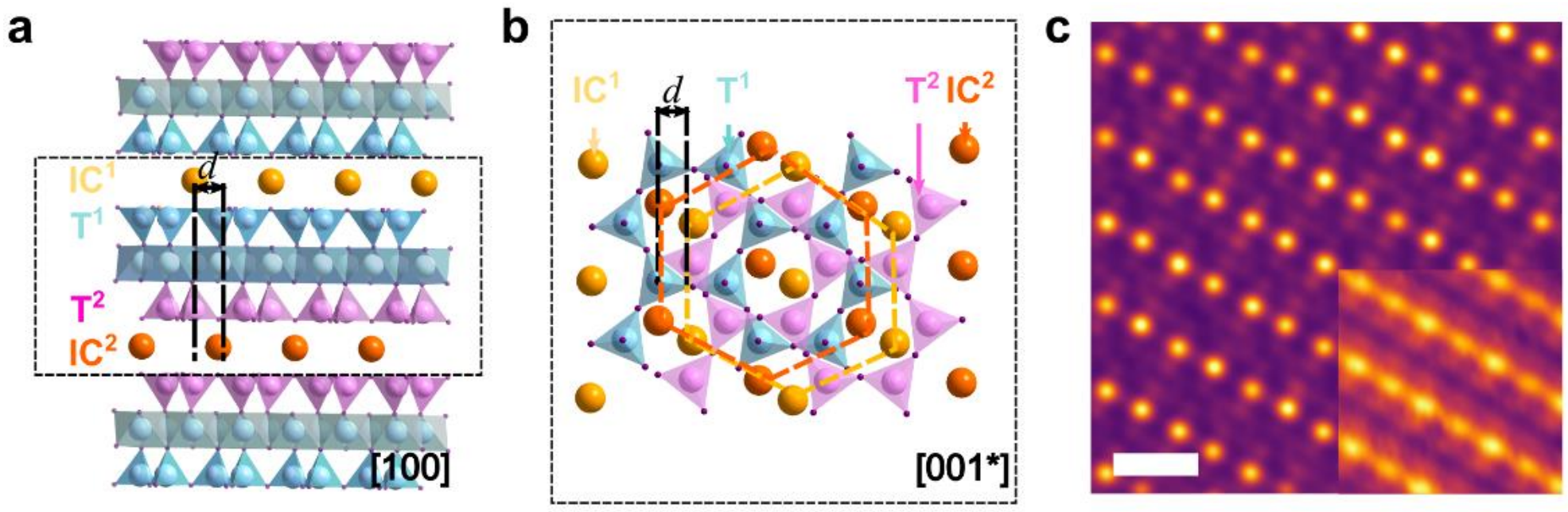

**Supplementary Figure S9| Stacking of interlayer cations in a perfect dehydrated mica/clay. a,** Cross-sectional atomic model of a trilayer mica. Tetrahedral (T) sublayers that are adjacent to a given interlayer cation plane are horizontally aligned with each other. This is illustrated by marking aligned T sub layers with the same colour (magenta, blue). The cations in adjacent interlayer spaces are marked in different colours (yellow, orange). **b**, Atomic model showing a plan view projection of clay. Only the $T^1$, $T^2$, $IC^1$ and $IC^2$ layers are shown for clarity, with *d* being the lateral displacement between the adjacent interlayer $Cs^+$. **c**, Simulated plan-view ADF-STEM image for Cs exchanged 3L vermiculite with inset being the experimental image from Fig. 2f. The zone axis is tilted ~3.8 degree away from [001]* along the ion-chain direction due to local specimen tilt.

In this section, we discuss the observed projected lateral arrangement for interlayer cations, as seen in Fig. 1f. Fig. S9a,b show cross-sectional and plan view atomic models for trilayer clay, to illustrate the expected interlayer-cation stacking behaviour. It can be seen that within each single TOT layer the 'T' sublayers (top $T^1$ and bottom $T^2$) are horizontally displaced with respect to each other, which results in a lateral displacement, *d,* between the cations ($IC^1$, $IC^2$) in nearest-neighbour interlayers. As shown in Fig. S9b,c, the projected linear patterns formed by $IC^1$, $IC^2$ matches the linear arrangement of interlayer ions observed in Fig. 1f (main text).

## 6. Measuring $Cs^+$ diffusivity in vermiculite flakes as a function of thickness

Ion exchange is considered to be purely a diffusion based phenomenon[13]. Indeed, the ion exchange speed is dominated by diffusion of ions in and out of the ion exchange site, rather than by a chemical reaction between the ions and the matrix[14]. Considering a one-dimensional model for interlayer ion diffusion[15], after time, *t*, the penetration distance, $\Delta P = (2Dt)^{0.5}$, and the diffusion constant, *D*, can be calculated as $<\Delta P^2>/2t$.

Previously, the diffusion constant, *D*, for different micas and clays has been estimated experimentally using crushed bulk powders[13,15], often by immersing the material in electrolyte and monitoring the chemical composition of the solution as a function of time. Note that such estimates provide an average over the entire crushed powder, which cannot distinguish the effects of crystal thickness, particle sizes and orientations in powder samples. In contrast, our plan view and cross-sectional STEM 'snapshots' obtained for different time points during the exchange process can be used to determine diffusivities as a function of crystal thickness, enabling direct observation of the penetration distance, $\Delta P$, of $Cs^+$ ions into the crystal at a particular interlayer. Examples for measurement of $\Delta P$ and the corresponding calculated *D* is shown in Table S3.1-3.15. This allows the calculation of the mean square penetration distance $<\Delta P^2> = \frac{\sum_{i=1}^{N-1} \Delta P_i^2}{N-1}$ ($N$ as the number of layers), and hence $D$ can be calculated.

It is of interest to compare our observed diffusivity values for thick samples with bulk literature data. Let us start with vermiculite. Depending on the particular type of vermiculite studied, and the type of ion being exchanged, literature reports for $D$ values in vermiculite bulk powders are in the range of $10^{-5}$ to 10 $\mu m^2\ s^{-1}$, with some specimens displaying diffusivity values too slow to be measured in a realistic time frame[16,17]. As a heavy ion with a high hydration enthalpy, $Cs^+$ diffuses orders of magnitude slower compared to ions with low hydration enthalpy like $Na^+$[8,17]. The diffusion constant we measured for $Cs^+$ ions in our multilayer vermiculite specimens is (7±3) x $10^{-3}$ $\mu m^2\ s^{-1}$ (Fig. 2d). This is consistent with the literature values of $10^{-2}$-$10^{-3}$ measured for $Cs^+$ by bulk powder experiments in vermiculite clays[16,17].

To measure the fast $Cs^+$ diffusivity in 2L and 3L samples, we isolated bilayer flakes with exceptionally large lateral dimensions (10- 35 µm, Fig. S10). This was necessary for accurate measurement of *D,* as smaller 2L and 3L flakes were completely exchanged within one second. In such cases, only lower bounds for $D$ can be provided (Table S3.3). To clarify how we conducted $Cs^+$ diffusivity $D_N$ calculations, the as-measured experimental layer-by-layer penetration distances $\Delta P_i$, from 15 vermiculite flakes of different thickness, are provided as Tables 3.1-3.15.

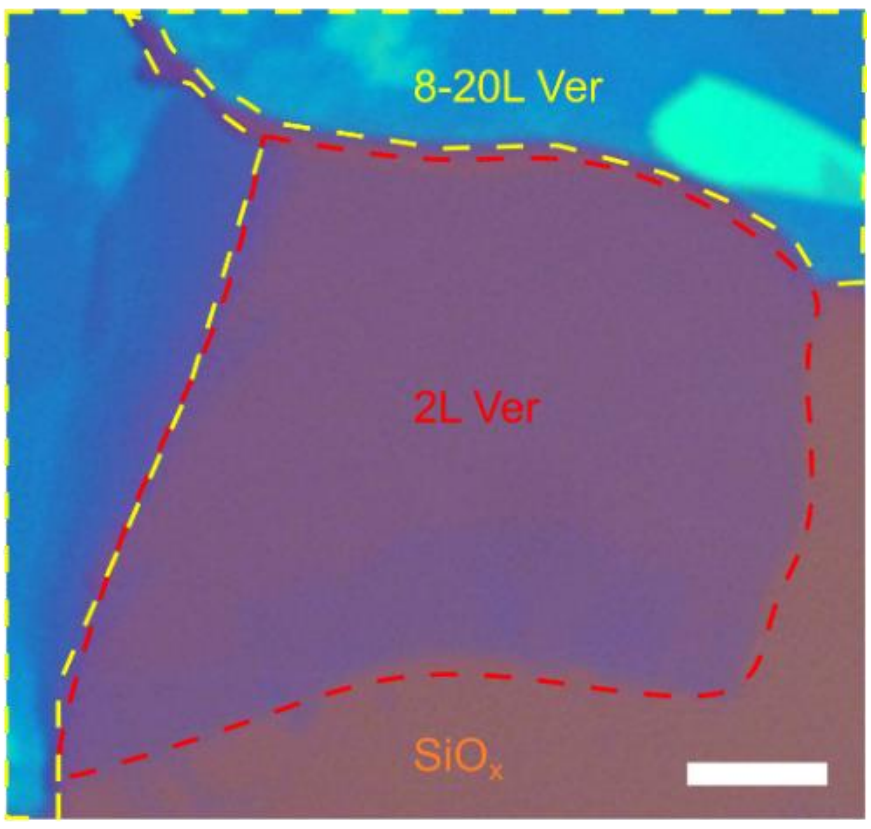

**Supplementary Figure S10| Example optical image of a 2L layer vermiculite flake used for $Cs^+$ diffusivity measurement. Data reported in Table S3.1.** The red dashed line shows the 2L vermiculite region (violet contrast). Scale bar, 10 µm.

**Table S3| Measured layer-by-layer penetration distances $\Delta P_i$ for $Cs^+$ diffusivity $D_N$ calculations, examples from 15 vermiculite flakes of different thickness with lateral sizes of 7 to 35 µm** (Summary data in Fig. 2d)

**Table S3.1|** A 2L vermiculite sample (lateral size of 2L region ≈35 µm, Fig. S10)

| Interlayer (*i*#) | $\Delta P_i$ (µm) | $\Delta P_i^2$ (µm$^2$) |
|---|---|---|
| 1 | 14.5 | 210.25 |
| Mean $\Delta P_i^2$, $<\Delta P^2>$ ($t$=1s) | | 210.25 |
| $D_{2L}$ | | 105.13 µm$^2$/s |

**Table S3.2|** A 2L vermiculite sample (lateral size of 2L region ≈21 µm)

| Interlayer (*i*#) | $\Delta P_i$ (µm) | $\Delta P_i^2$ (µm$^2$) |
|---|---|---|
| 1 | 11.0 | 121 |
| Mean $\Delta P_i^2$, $<\Delta P^2>$ ($t$=1s) | | 121 |
| $D_{2L}$ | | 60.5 µm$^2$/s |

**Table S3.3|** A 2L vermiculite sample (lateral size of 2L region ≈7 µm)

| Interlayer (*i*#) | $\Delta P_i$ (µm) | $\Delta P_i^2$ (µm$^2$) |
|---|---|---|
| 1 | 7.0 | 49 |
| Mean $\Delta P_i^2$, $<\Delta P^2>$ ($t$=1s) | | 49 |
| $D_{2L}$ | | 24.5 µm$^2$/s |

**Table S3.4|** A 3L vermiculite sample (lateral size of 3L region ≈20 µm)

| Interlayer (*i*#) | $\Delta P_i$ (µm) | $\Delta P_i^2$ (µm$^2$) |
|---|---|---|
| 1 | 9.0 | 81 |
| 2 | 6.8 | 46.24 |
| Mean $\Delta P_i^2$, $<\Delta P^2>$ ($t$=1s) | | 63.62 |
| $D_{3L}$ | | 31.81 µm$^2$/s |

**Table S3.5|** A 3L vermiculite sample (lateral size of 3L region ≈10 µm)

| Interlayer (*i*#) | $\Delta P_i$ (µm) | $\Delta P_i^2$ (µm$^2$) |
|---|---|---|
| 1 | 4.00 | 16 |
| 2 | 7.00 | 49 |
| Mean $\Delta P_i^2$, $<\Delta P^2>$ ($t$=1s) | | 32.5 |
| $\boldsymbol{D_{3L}}$ | | 16.25 $\mu m^2/s$ |

**Table S3.6 |** A 3L vermiculite sample (lateral size of 3L region ≈8 µm)

| Interlayer (*i*#) | $\Delta P_i$ (µm) | $\Delta P_i^2$ ($\mu m^2$) |
|---|---|---|
| 1 | 5.00 | 25 |
| 2 | 4.00 | 16 |
| Mean $\Delta P_i^2$, $<\Delta P^2>$ ($t$=1s) | | 20.5 |
| $\boldsymbol{D_{3L}}$ | | 10.25 $\mu m^2/s$ |

**Table S3.7|** A 3L vermiculite sample (lateral size of 3L region ≈7 µm)

| Interlayer (*i*#) | $\Delta P_i$ (µm) | $\Delta P_i^2$ ($\mu m^2$) |
|---|---|---|
| 2 | 2.00 | 4 |
| 2 | 1.70 | 2.89 |
| Mean $\Delta P_i^2$, $<\Delta P^2>$ ($t$=1s) | | 9.45 |
| $\boldsymbol{D_{3L}}$ | | 3.45 $\mu m^2/s$ |

**Table S3.8|** A 4L vermiculite sample (lateral size of 4L region ≈13 µm)

| Interlayer (*i*#) | $\Delta P_i$ (µm) | $\Delta P_i^2$ ($\mu m^2$) |
|---|---|---|
| 1 | 6.00 | 36 |
| 2 | 1.00 | 1 |
| 3 | 3.00 | 9 |
| Mean $\Delta P_i^2$, $<\Delta P^2>$ ($t$=1s) | | 15.33 |
| $\boldsymbol{D_{4L}}$ | | 7.66 $\mu m^2/s$ |

**Table S3.9|** A 5L vermiculite sample (lateral size of 5L region ≈10 µm)

| Interlayer (i#) | $\Delta P_i$ (µm) | $\Delta P_i^2$ ($\mu m^2$) |
|---|---|---|
| 1 | 2.20 | 4.84 |
| 2 | 1.50 | 2.25 |
| 3 | 0.50 | 0.25 |
| 4 | 2.20 | 4.84 |
| Mean $\Delta P_i^2$, $<\Delta P^2>$ ($t$=10s) | | 3.05 |
| $D_{5L}$ | | 0.15 μm²/s |

**Table S3.10|** A 5L vermiculite sample (lateral size of 5L region ≈15 μm)

| Interlayer (i#) | $\Delta P_i$ (μm) | $\Delta P_i^2$ (μm²) |
|---|---|---|
| 1 | 0.50 | 0.25 |
| 2 | 0.34 | 0.1156 |
| 3 | 0.30 | 0.09 |
| 4 | 2.05 | 4.2 |
| Mean $\Delta P_i^2$, $<\Delta P^2>$ ($t$=1s) | | 1.16 |
| $D_{5L}$ | | 0.58 μm²/s |

**Table S3.11|** A 7L vermiculite sample (lateral size of 7L region ≈7 μm)

| Interlayer (i#) | $\Delta P_i$ (μm) | $\Delta P_i^2$ (μm²) |
|---|---|---|
| 1 | 1.00 | 1 |
| 2 | 0.30 | 0.09 |
| 3 | 0.10 | 0.01 |
| 4 | 0.20 | 0.04 |
| 5 | 0.30 | 0.09 |
| 6 | 0.60 | 0.36 |
| Mean $\Delta P_i^2$, $<\Delta P^2>$ ($t$=1s) | | 0.27 |
| $D_{7L}$ | | 0.14 μm²/s |

**Table S3.12|** A 12L vermiculite sample (lateral size of 7L region ≈12 μm)

| Interlayer (i#) | $\Delta P_i$ (μm) | $\Delta P_i^2$ (μm²) |
|---|---|---|
| 1 | 0.40 | 0.16 |
| 2 | 0.35 | 0.1225 |
| 3 | 0.10 | 0.01 |
| 4 | 0.60 | 0.36 |
| 5 | 0.00 | 0 |
| 6 | 0.60 | 0.36 |
| 7 | 0.10 | 0.01 |
| 8 | 0.40 | 0.16 |
| 9 | 0.06 | 0.0036 |
| 10 | 0.10 | 0.01 |
| 11 | 0.40 | 0.16 |
| Mean $\Delta P_i^2$, $<\Delta P^2>$ ($t$=10s) | | 0.12 |
| $\boldsymbol{D_{12L}}$ | | 0.006 μm²/s |

**Table S3.13|** A 33L vermiculite sample (lateral size of 33L region ≈14 μm)

| Interlayer (i#) | $\Delta P_i$ (μm) | $\Delta P_i^2$ (μm$^2$) |
|---|---|---|
| 1 | 0.24 | 0.0576 |
| 2 | 0.26 | 0.0676 |
| 3 | 0.00 | 0 |
| 4 | 0.24 | 0.0576 |
| 5 | 0.22 | 0.0484 |
| 6 | 0.00 | 0 |
| 7 | 0.23 | 0.0529 |
| 8 | 0.20 | 0.04 |
| 9 | 0.28 | 0.0784 |
| 10 | 0.00 | 0 |
| 11 | 0.25 | 0.0625 |
| 12 | 0.50 | 0.25 |
| 13 | 0.32 | 0.1 |
| 14 | 0.22 | 0.0484 |
| 15 | 0.00 | 0 |
| 16 | 0.24 | 0.0576 |
| 17 | 0.05 | 0.0025 |
| 18 | 0.10 | 0.01 |
| 19 | 0.00 | 0 |
| 20 | 0.00 | 0 |
| 21 | 0.34 | 0.1156 |
| 22 | 0.51 | 0.2601 |
| 23 | 0.00 | 0 |
| 24 | 0.12 | 0.0144 |
| 25 | 0.12 | 0.0144 |
| 26 | 0.10 | 0.01 |
| 27 | 0.10 | 0.01 |
| 28 | 0.15 | 0.0225 |
| 29 | 0.10 | 0.01 |
| 30 | 0.00 | 0 |
| 31 | 0.42 | 0.1764 |
| 32 | 0.25 | 0.0625 |
| Mean $\Delta P_i^2$, $<\Delta P^2>$ ($t$=10s) | | 0.05 |
| $\boldsymbol{D_{33L}}$ | | 0.003 µm$^2$/s |

**Table S3.14|** A 42L vermiculite sample (lateral size of 42L region ≈9 µm)

| Interlayer (i#) | $\Delta P_i$ (µm) | $\Delta P_i^2$ (µm$^2$) |
|---|---|---|
| 1 | 0.20 | 0.04 |
| 2 | 0.10 | 0.01 |
| 3 | 0.50 | 0.25 |
| 4 | 0.08 | 0.0064 |
| 5 | 0.3 | 0.09 |
| 6 | 0.36 | 0.1296 |
| 7 | 0.13 | 0.0169 |
| 8 | 0.60 | 0.36 |
| 9 | 0.10 | 0.01 |
| 10 | 0.40 | 0.16 |
| 11 | 0.30 | 0.09 |
| 12 | 0.30 | 0.09 |
| 13 | 0.03 | 0.0009 |
| 14 | 0.45 | 0.2025 |
| 15 | 0.30 | 0.09 |
| 16 | 0.28 | 0.0784 |
| 17 | 0.30 | 0.09 |
| 18 | 0.25 | 0.0625 |
| 19 | 0.10 | 0.01 |
| 20 | 0.30 | 0.09 |
| 21 | 0.25 | 0.0625 |
| 22 | 0.20 | 0.04 |
| 23 | 0.10 | 0.01 |
| 24 | 0.50 | 0.25 |
| 25 | 0.09 | 0.0081 |
| 26 | 0.64 | 0.4096 |
| 27 | 0.39 | 0.1521 |
| 28 | 0.30 | 0.09 |
| 29 | 0.22 | 0.0484 |
| 30 | 0.20 | 0.04 |
| 31 | 0.30 | 0.09 |
| 32 | 0.24 | 0.0576 |
| 33 | 0.14 | 0.0196 |
| 34 | 0.12 | 0.0144 |
| 35 | 0.30 | 0.09 |
| 36 | 0.10 | 0.01 |
| 37 | 0.05 | 0.0025 |
| 38 | 0.60 | 0.36 |
| 39 | 0.00 | 0 |
| 40 | 0.40 | 0.16 |
| 41 | 0.60 | 0.36 |
| Mean $\Delta P_i^2$, $<\Delta P^2>$ ($t$=10s) | | 0.10 |
| $\boldsymbol{D_{42L}}$ | | 0.005 $\mu m^2/s$ |

**Table S3.15|** A 66L vermiculite sample (lateral size of 66L region ≈24 µm)

| Interlayer (i#) | $\Delta P_i$ (µm) | $\Delta P_i^2$ ($\mu m^2$) |
|---|---|---|
| 1 | 0.40 | 0.16 |
| 2 | 0.30 | 0.09 |
| 3 | 0.70 | 0.49 |
| 4 | 0.00 | 0 |
| 5 | 0.00 | 0 |
| 6 | 0.30 | 0.09 |
| 7 | 0.20 | 0.04 |
| 8 | 0.45 | 0.2025 |
| 9 | 0.25 | 0.0625 |
| 10 | 0.00 | 0 |
| 11 | 0.74 | 0.5476 |
| 12 | 0.18 | 0.0324 |
| 13 | 0.14 | 0.0196 |
| 14 | 0.00 | 0 |
| 15 | 0.00 | 0 |
| 16 | 0.72 | 0.5184 |
| 17 | 0.21 | 0.0441 |
| 18 | 0.30 | 0.09 |
| 19 | 0.75 | 0.5625 |
| 20 | 0.03 | 0.0009 |
| 21 | 0.35 | 0.1225 |
| 22 | 0.20 | 0.04 |
| 23 | 0.60 | 0.36 |
| 24 | 0.20 | 0.04 |
| 25 | 0.60 | 0.36 |
| 26 | 0.00 | 0 |
| 27 | 0.06 | 0.0036 |
| 28 | 0.00 | 0 |
| 29 | 0.06 | 0.0036 |
| 30 | 0.00 | 0 |
| 31 | 0.10 | 0.01 |
| 32 | 0.60 | 0.36 |
| 33 | 0.10 | 0.01 |
| 34 | 0.70 | 0.49 |
| 35 | 0.00 | 0 |
| 36 | 0.06 | 0.0036 |
| 37 | 0.20 | 0.04 |
| 38 | 0.66 | 0.4356 |
| 39 | 0.10 | 0.01 |
| 40 | 0.60 | 0.36 |
| 41 | 0.10 | 0.01 |
| 42 | 0.06 | 0.0036 |
| 43 | 0.06 | 0.0036 |
| 44 | 0.00 | 0 |
| 45 | 0.00 | 0 |
| 46 | 0.30 | 0.09 |
| 47 | 0.50 | 0.25 |
| 48 | 0.00 | 0 |
| 49 | 0.50 | 0.25 |
| 50 | 0.10 | 0.01 |
| 51 | 0.50 | 0.25 |
| 52 | 0.00 | 0 |
| 53 | 0.60 | 0.36 |
| 54 | 0.10 | 0.01 |
| 55 | 0.10 | 0.01 |
| 56 | 0.60 | 0.36 |
| 57 | 0.20 | 0.04 |
| 58 | 0.80 | 0.64 |
| 59 | 0.15 | 0.0225 |
| 60 | 0.20 | 0.04 |
| 61 | 0.00 | 0 |
| 62 | 0.70 | 0.49 |
| 63 | 0.10 | 0.01 |
| 64 | 0.70 | 0.49 |
| 65 | 0.10 | 0.01 |
| 60 | 0.20 | 0.04 |
| 61 | 0.00 | 0 |
| 62 | 0.70 | 0.49 |
| 63 | 0.10 | 0.01 |
| 64 | 0.70 | 0.49 |
| 65 | 0.10 | 0.01 |
| 63 | 0.10 | 0.01 |
| Mean $\Delta P_i^2$, $<\Delta P^2>$ ($t$=10s) | | 0.14 |
| $\boldsymbol{D_{42L}}$ | | 0.007 μm$^2$/s |

## 7. Enhancing $Cs^+$ exchange speed using exfoliated vermiculite laminate

Exfoliated vermiculite laminates were prepared using the method reported in previous work[18]. In brief, 100 mg of bulk Mg vermiculite is mixed with a saturated solution of NaCl (3 g/ml) and heated under reflux conditions for 24 hrs. The vermiculite is then filtered and washed thoroughly using DI water and mixed with 2 M LiCl under reflux conditions for 24 hrs. The vermiculite is then filtered and washed thoroughly using DI water to remove any residual salt. Vermiculite crystals after washing are dispersed in 50 ml DI water and sonicated for a period of 30 min to facilitate exfoliation. The solution thus obtained is centrifuged at 4000 rpm until no sediments are observed. The supernatant is collected and used for further experiments. Few layer vermiculite laminates are obtained using the above-mentioned method consisting of ~30% bilayer and ~15% trilayer crystallites (Li-vermiculite, thickness distribution shown in Fig. S11c). These are then filtered through porous alumina substrate, using vacuum-assisted self-assembly to create a laminate membrane. The 2D laminate membranes thus obtained are dried and the interlayer cations are exchanged with $Mg^{2+}$ ions by treating with 2M $MgCl_2$ for a period of 3 hrs (Mg-vermiculite membranes, see inset in Fig. S11c). The membranes are dried and used for further ion exchange experiments. For comparison, similar ‘bulk’ samples were also made from an equal quantity of bulk Mg vermiculite powder.

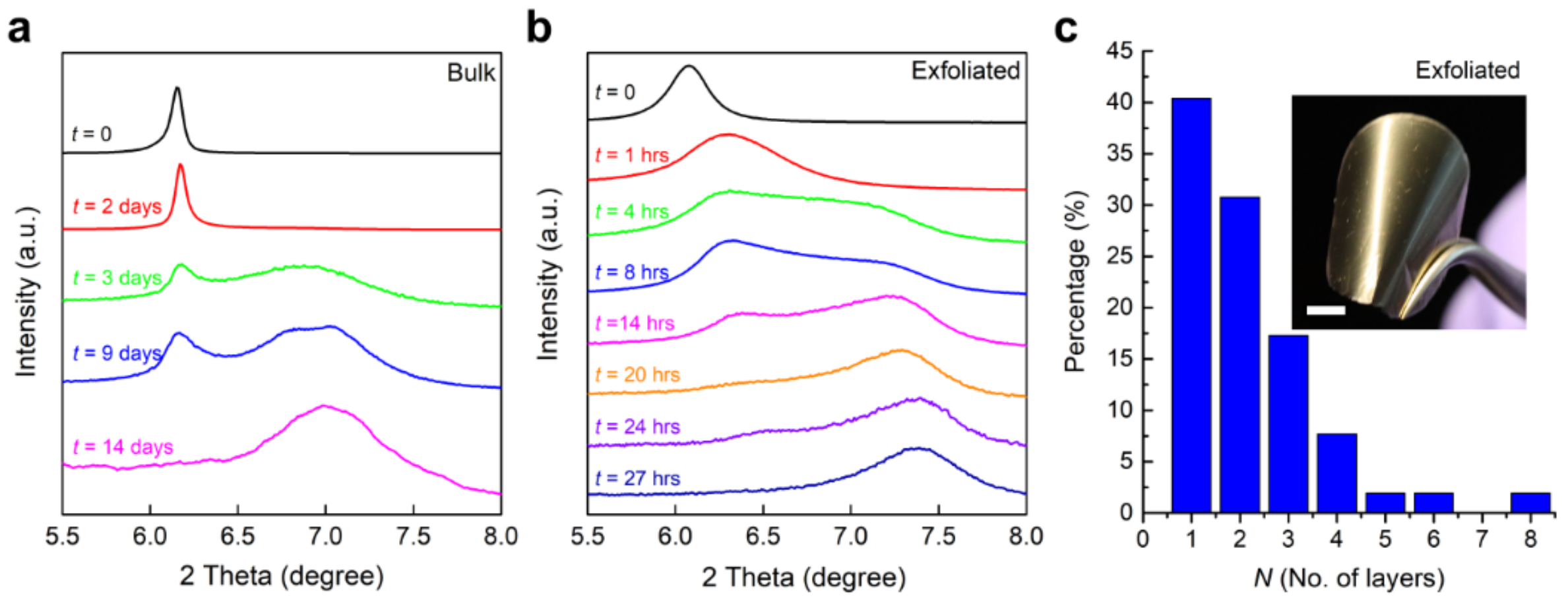

**Supplementary Figure S11| Exchange speed characterization of vermiculite samples with different exposure times for $Cs^+$ ion exchange, analysed using XRD. a,** XRD taken from bulk crystal (*t* being the CsCl solution treatment time). **b,** XRD taken from the laminates made from exfoliated nanosheets. A right shift of (002) peaks can be observed with increasing exposure time due to the contraction of interlayer spacing caused by $Cs^+$ incorporation. **c,** Flake thickness distribution in the laminates measured by AFM. Inset, optical image of a typical Mg-vermiculite laminate (membrane). Scale bar, 5 mm.

For ion exchange experiments, the as-fabricated membranes (inset of Fig. S11c) are cut into several samples of the same lateral size (~5 mm x 5 mm). Both the exfoliated and 'bulk' membrane samples are then exposed to 1 M CsCl for a specified period, followed by an extensive wash with DI water and immersion in 100 ml of DI water for 2 days to remove any excess ions. The exchanged samples are then dried at 50 °C for 24 hrs prior to XRD analysis. XRD data of both the exfoliated membrane and the bulk samples were collected using a Rigaku smart lab thin film XRD system (Cu-Kα radiation) operated at 1.8 kW.

To characterise the exchange process, we note that $Cs^+$ exchange reduces the interlayer spacing of vermiculite from ~1.4 nm down to ~1.1 nm (see Fig. S5g). This allows for the monitoring of the exchange process with XRD diffraction analysis. We found that it takes 2-3 days to observe any detectable Cs exchange in the bulk crystal (Fig. S11a). In stark contrast, for exfoliated laminates, we observe significant exchange within 1 hour – a factor of ~100 faster (Fig. S11b). Following this trend, we find that it takes ~14 days for bulk crystals to stabilise at ~1.26 nm interlayer distance. In contrast, exfoliated laminates reach this value in less than a day (factor of ~10 faster) and, unlike the bulk crystal, stabilise at a smaller interlayer distance ~1.19 nm that corresponds to virtually complete $Cs^+$ exchange. These data demonstrate that ion exchange in exfoliated vermiculite laminates is 1-2 orders of magnitude faster than in the bulk crystals. While this is slower than the 2-4 orders of magnitude factor observed in monocrystalline flakes, the difference is hardly surprising. First, the laminates contain a significant (~30%) proportion of >3 layer nanosheets (Fig. S11c), in which $Cs^+$ diffusion is expected to be >10 times slower than for bilayer crystals. Second, due to their macroscopic (millimetre) size, ion diffusion within the laminate itself should contribute to the measured diffusion constant. Note that these factors do not arise in our highly controlled single-crystalline STEM samples. Despite this quantitative difference, this experiment confirms our observation that ion exchange is orders of magnitude faster in atomically thin clays compared to bulk crystals.

## 8. Estimation of $Cs^+$ diffusivity in biotite and muscovite

Snapshot ADF-STEM diffusivity measurements were also performed on few layer specimens of muscovite and biotite for various thicknesses (up to 9 months)[14]. For both muscovite and biotite, only a few nanometres close to the edges of the crystal were found to have interlayer exchange providing maximum diffusivity values for few-layer crystals of just $10^{-11}$ µm$^2$ s$^{-1}$ ($D_{6L\text{-}Mus}$ ~$10^{-11}$ µm$^2$ s$^{-1}$, and $D_{4L\text{-}Bio}$ ~$10^{-11}$ µm$^2$ s$^{-1}$, Fig. S12). Literature studies have also reported extremely slow interlayer ion exchange in muscovite; below what is measurable in a realistic time frame[15,19].

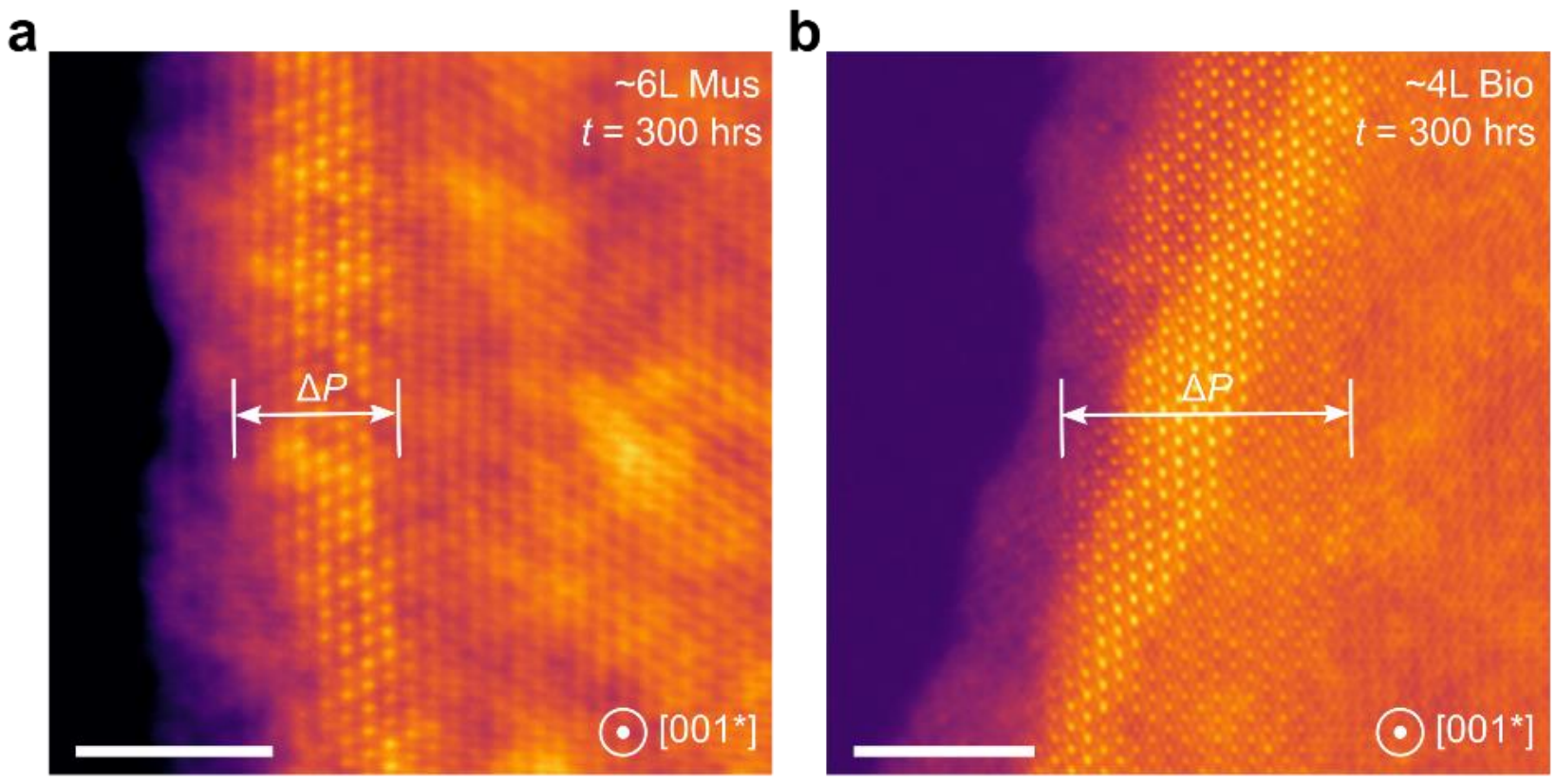

**Supplementary Figure S12|** Plan view ADF-STEM images of **a,** few-layered muscovite (Mus), **b,** Few-layered biotite (Bio), both Cs-exchanged for 300 hours. $Cs^+$ ions are visible as bright yellow dots. The largest penetration depth of Cs into the interlayer space in the crystal (Δ*P*) is indicated, yielding $D_{6L\text{-}Mus}$ ~$10^{-11}$ µm$^2$ s$^{-1}$, and $D_{4L\text{-}Bio}$ ~$10^{-11}$ µm$^2$ s$^{-1}$. Scale bars, 5 nm.

The intrinsic chemical difference between biotite (muscovite) and vermiculite can explain their distinct differences in interlayer ion diffusivity. Muscovite and biotite are micas that have a higher negative structural charge (1 $e^-$ per structural formula, unit shown in Table S2) than vermiculite clay (0.6-0.9 $e^-$ per structural formula unit)[20]. This impacts interlayer ion diffusivity in two ways. First, the higher negative structural charge of muscovite and biotite indicate a greater density of pristine interlayer cations for charge compensation, and therefore there are fewer interlayer vacancy sites to stimulate the ion exchange process[21]. Second, the backbone TOT layers are more tightly bound with the interlayer cations as they possess higher charge density, leading to slightly lower spacing as observed in Fig. S3. The energy barrier for interlayer expansion required for $Cs^+$ migration should then be significantly higher in biotite (muscovite) compared to vermiculite.

An advantage of working with these slow exchanging crystals is that imaging them provides ‘snapshots’ of the earliest stages of ion penetration into the interlayer galleries. Fig. S12 shows that ions were uniformly distributed along the edges and were never concentrated along defects. Nor did they show any preference for diffusion along a particular crystallographic direction within the basal interlayer space. This further shows that ion exchange in our samples is the result of uniform ion penetration into the interlayer space through the entire length of exposed crystal edges.

## 9. Observing fast ion exchange at the interfaces of restacked biotite flakes

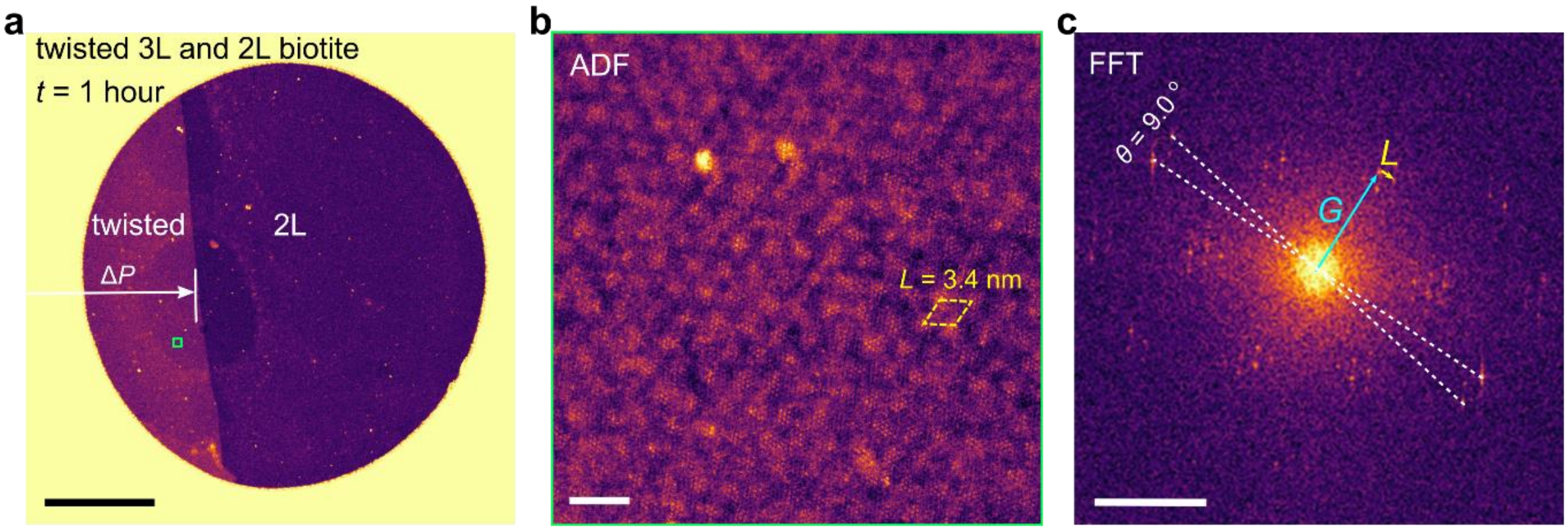

**Supplementary Figure S13| Plan view characterization for ion exchange at the interface of twisted few-layer biotite flakes. a,** Low-magnification ADF-STEM image taken from a Cs-biotite sample, yellow border area is the TEM support grid. The right-side is a 2L region where no $Cs^+$ was observed, whereas the left side is a 9.0° twisted region, found to have been unexpectedly exchanged by $Cs^+$ over a micron-sized scale region ($\Delta P$ > 2µm). Scale bar, 2 µm. **b**, Representative zoomed-in ADF-STEM image taken from the local region marked by a green square in **a**, showing the formation of $Cs^+$ islands with a periodicity of ~3.4 nm. Scale bar, 5 nm. **c**, Fast Fourier transform (FFT) of **b**, where the $Cs^+$ lattice/biotite matrix spots (*G* corresponding to a lattice vector of 5.3 Å), are surrounded by satellites caused by the periodically occurring $Cs^+$ islands (*L* corresponding to a $Cs^+$ superlattice vector of 3.4 nm), showing the exchanged $Cs^+$ ions are spatially modulated by the twist angle. Scale bar, 2 $nm^{-1}$.

Fig. S13a shows plan-view characterization results for a biotite flake that consists of both a restacked region and a bilayer region. The restacked region ($\Delta P$ > 2 µm, see Fig. S13a) was found to have been fully exchanged by $Cs^+$ after $t$= 1 hour, which gives $D_{\text{twist-Bio}} > 10^{-4}$ µm$^2$ s$^{-1}$, whereas the 2L region shows no Cs exchange. The high-resolution ADF-STEM image of the exchanged twisted region shows periodic bright contrast generated by $Cs^+$ islands, consistent with the other examples shown in Fig. 3, main text. Such contrast is not generated by the moiré interference from the aluminosilicate layers and requires the addition of $Cs^+$ islands to achieve a good agreement with image simulation results (Fig. S14).

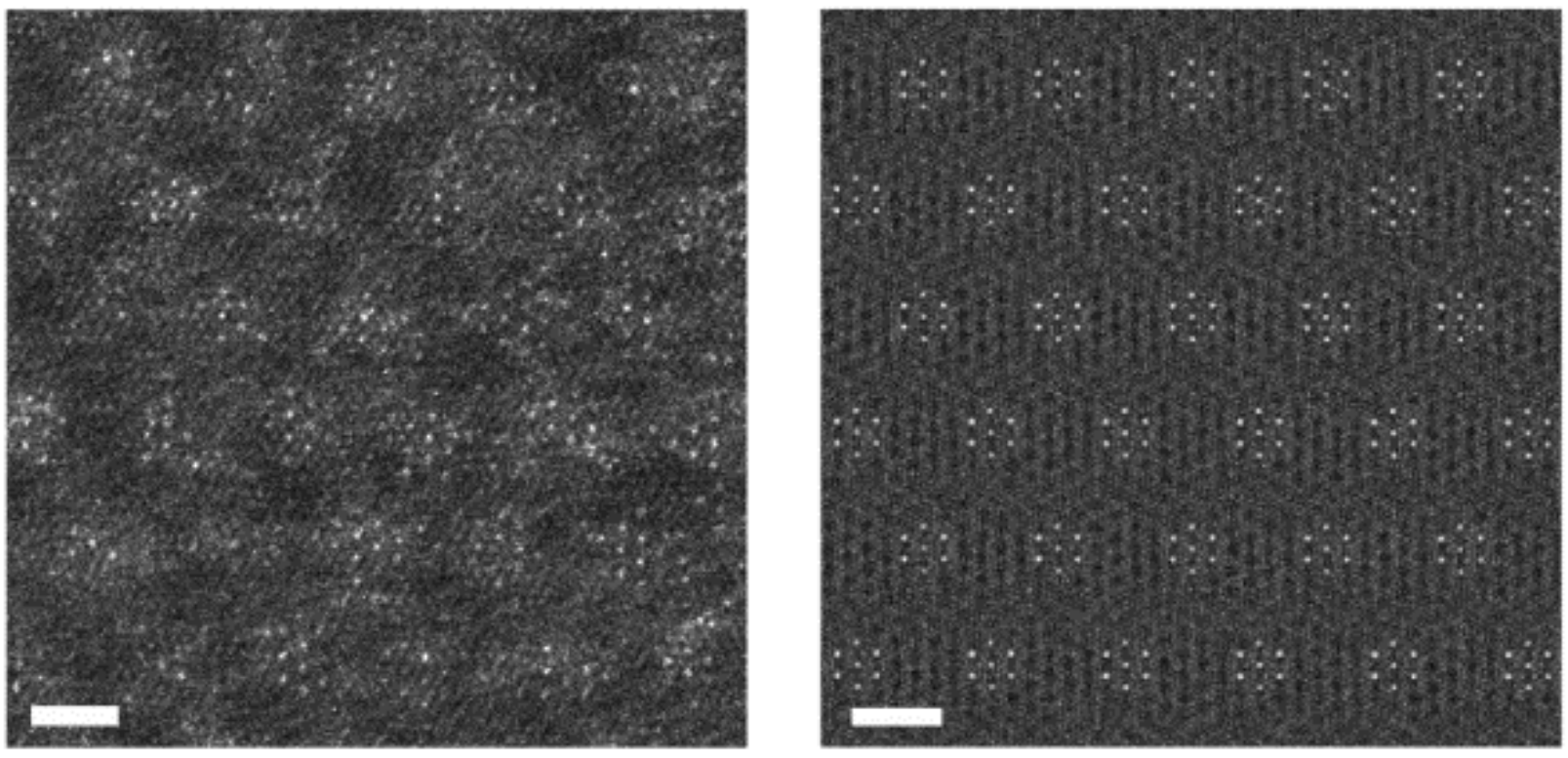

**Supplementary Figure S14| Comparison between experimental image and simulations for Cs island exchanged 9.5° twisted bilayer biotite flakes.** Left panel is the ADF-STEM experimental image. Right panel is the simulated image, where Poisson noise with λ = 3 and signal-to-noise ratio

of ~26:1 was included, using the flag generation plugin for Digital Micrograph developed by Braidy, et al[22]. Scale bar, 2 nm.

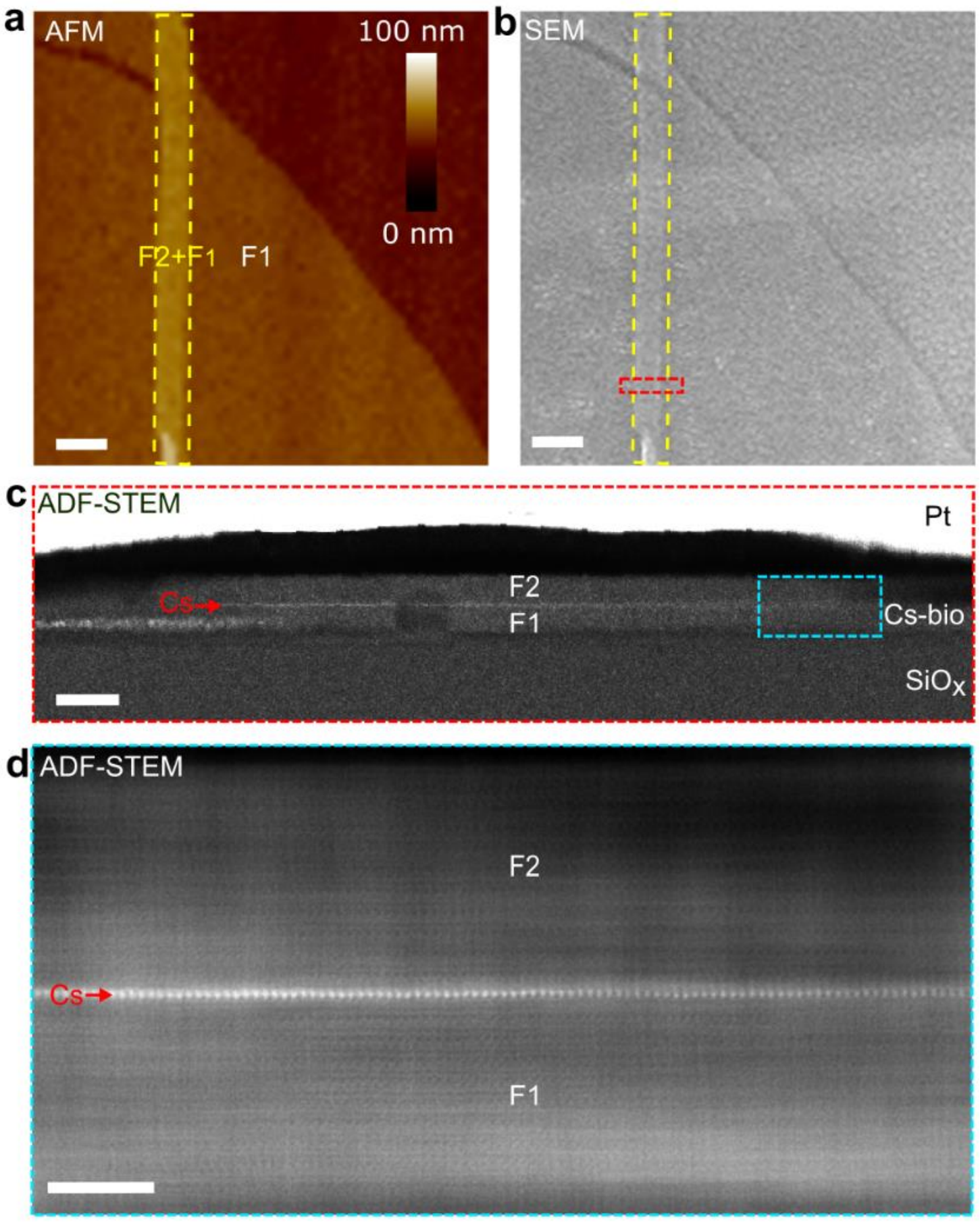

**Supplementary Figure S15| Ion exchange at the interface of twisted multilayer biotite flakes. a,** AFM topography image taken from a biotite sample consisting of two restacked flakes. The bright-contrast region shows the area where the two biotite flakes (labelled as F1, and F2) are restacked. Scale bar, 500 nm**. b,** Scanning electron micrograph of the sample in **a**. The red marker shows the region that was lifted out for FIB TEM cross-sectioning. Scale bar, 500 nm. **c,** Corresponding ADF-STEM image taken from the as-prepared TEM cross-sectional specimen. Scale bar, 20 nm. **d,** Magnified ADF-STEM image taken from the area marked by the blue rectangle in **b**. Scale bar, 5 nm.

Fig. S15 shows cross-sectional characterization for two restacked multilayer biotite flakes, that were exposed to Cs ion exchange for *t*= 1 hour after being restacked. Fig. S15c,d show high-resolution images indicating that the interlayer space between the two misaligned crystals indeed hosts only one plane of $Cs^+$ ions, visible as a row of bright dots. In contrast, the interlayer spaces in each of the original flakes show no Cs exchange at all, as expected from the slow ion exchange of biotite. From performing many similar STEM studies of ion exchange for the interface between twisted biotite crystals, we find that ion exchange consistently takes place within ~1 hour – an apparent enhancement of ~7 orders of magnitude faster compared to the interlayer space in biotite crystals (Fig. S12b, S13a). This is the case even if the two twisted flakes are relatively thick (≳10 layers thick). We also note that the interlayer distance at the interface between the two flakes is just ~1.0 nm. This rules out the presence of contamination, like hydrocarbons, at the interface as a contributor to fast ion exchange. Instead, it is consistent with our explanation outlined in the main text, which attributes the fast ion exchange to the weaker interlayer binding energy between the misaligned flakes (Fig. S21).

### 10. AFM characterization for swelling behaviour of vermiculite flakes

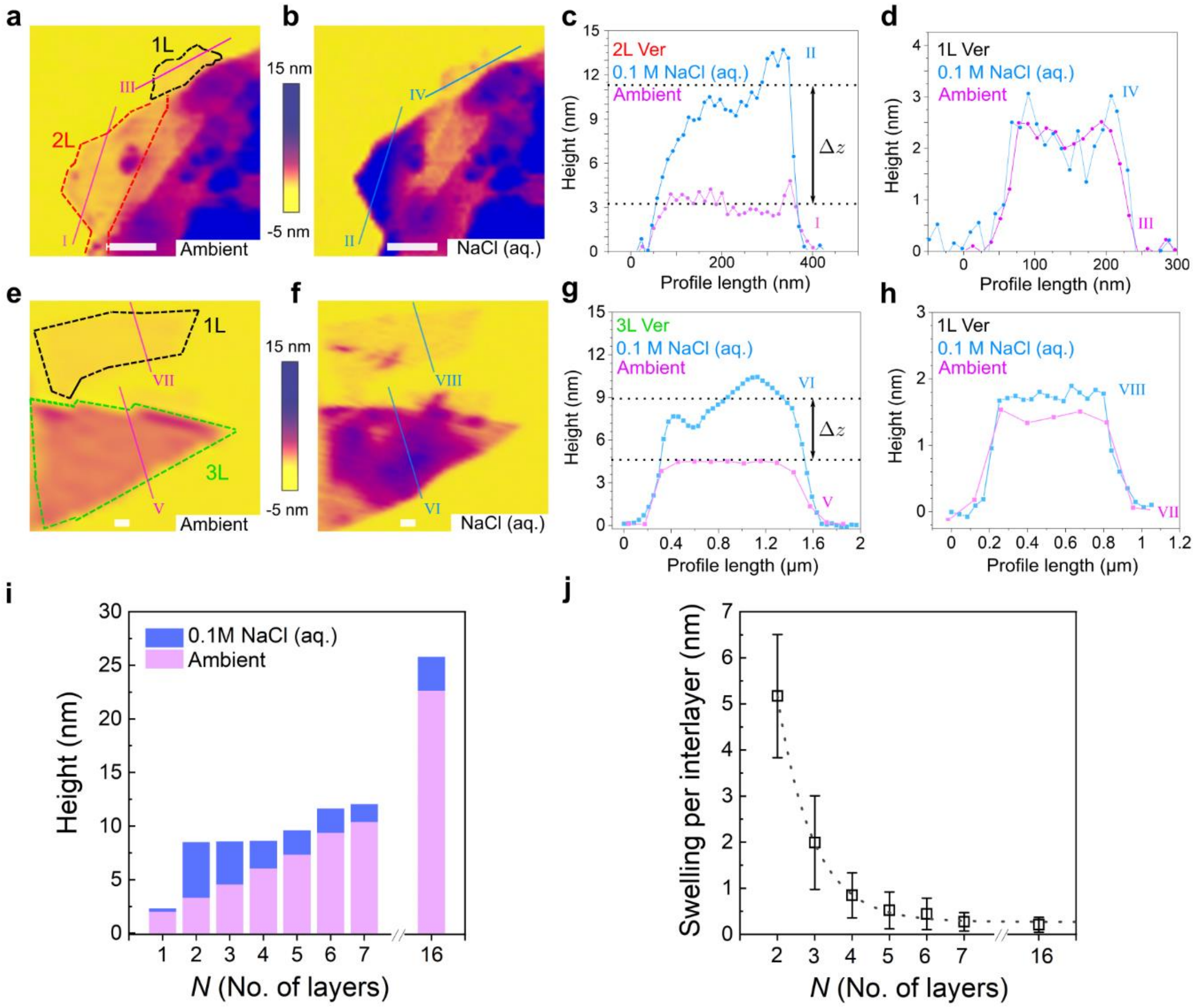

**Supplementary Figure S16| Measurements comparing swelling behaviour of vermiculite with different thicknesses in NaCl solution. a-d,** AFM results for a 2L vermiculite region (outlined by red dashed lines) and a local 1L region (outlined by black dashed lines). **a, b,** AFM height image taken **a,** in ambient air and **b,** in aqueous 0.1 M NaCl. **c,d,** Corresponding height profiles in air and liquid extracted from the positions of the lines (I, II, III, IV) for **c,** 2L region and **d,** 1L region. **e-h,** AFM results for a 3L vermiculite (green dashed lines) containing a local 1L region (black dashed lines). **e, f,** AFM height image taken **e,** in ambient air, and **f,** in aqueous 0.1 M NaCl. **g,h,** Corresponding height profiles extracted from the positions of the lines (V, VI, VII, VIII) for **g,** 3L region, and **h,** 1L region. **i,** Statistical results showing the AFM measured height ($z$) values for vermiculite as a function of $N$, in ambient air (magenta) and liquid NaCl (blue). Swelling values ($\Delta z$) are extracted as height difference in liquid and ambient, with the standard deviation extracted from at least three different measurements. **j,** Mean swelling values per interlayer as a function of $N$, calculated as $\Delta z \times (N\text{-}1)^{-1}$ using $\Delta z$ values shown in **i**. Dotted line is a guide to the eye. Scale bars, 200 nm.

To investigate the possible effect of crystal thickness on interlayer expandability, we measured the height profile of vermiculite flakes in ambient conditions and in liquid using AFM (Fig. S16). Swelling values for a certain flake are extracted from the difference in height measured in ambient air and in liquid environment. To ensure that the difference in height is not due to water intercalation between the vermiculite flake and the substrate, we modified our sample preparation with respect to the recipe described in Section 2. Prior to flake exfoliation, the $SiO_x$/Si substrates were coated with an adhesion layer of SU8 3005 epoxy and then photo-cured under a UV source. Vermiculite samples were then mechanically exfoliated onto the as-prepared substrates. The samples are 'hard baked', typically at 60 °C for 10 minutes, to further promote polymer/clay adhesion. For measurements in ambient air, AFM scans were taken in ambient conditions in a 1000 class clean room environment at ~50% relative humidity and ~21 °C. For the measurement of flakes immersed in liquid, we used a 0.1M NaCl aqueous solution.

Fig. S16 shows data for flakes of different thicknesses ranging from 2-16 layers. Statistically, we found that there is a distinct difference in swelling behaviour between the 2L vermiculite and thick crystals (i.e. ≥ 7 layer). Bilayer vermiculite flakes swell by ~5 nm per interlayer, compared to <0.5 nm for flakes with $N \geq 7$ layers. Moreover, the thickness-normalised interlayer expansion decayed rapidly with the number of layers, stabilising for flakes thicker than ~7 layers (Fig. S16). To confirm that the increase in height for samples measured in liquid was indeed due to interlayer expansion, we measured monolayer (1L) regions on the same flakes as a reference (Fig. S16d,h). These monolayer regions present negligible height increase in liquid, as expected since there is no interlayer space. The enhanced interlayer expandability in 2L and 3L vermiculite is expected to be the main factor explaining the strong dependence of ion diffusion on flake thickness.

## 11. Identifying the $Cs^+$ ions adsorbed on mica and clay surfaces

By encapsulating the mica and clay flakes between graphite and a $SiO_x$/Si substrate, we find that it is possible to preserve adsorbed surface ions (Fig. S17). Cross-sectional STEM allows for the measurement of the out-of-plane separation, $z$, between the surface-ion plane and the outmost edge of the 'T' aluminosilicate layer at the basal surface[23,24]. We find a relatively large variability in $z$ for surface adsorbed $Cs^+$ ions. For different specimens we observe values ranging from $z \approx 1.8 \pm 0.2$ Å to $z \approx 2.5 \pm 0.6$ Å. We speculate that this large variability in $z$ for $Cs^+$ is due to the complex hydration structures possible for surface $Cs^+$ ions[23-27]. Although STEM imaging has not previously been demonstrated for adsorbed surface ions, different experimental techniques have found that different $Cs^+$ hydration complexes yield different $z$ values[19,24,26]. While, in principle, the formation of these complexes could also be related to the local chemistry environment at mica(clay) and $SiO_x$/graphite interfaces, we did not find a preference for $Cs^+$ to occupy a specific interface for our samples (either with graphite or silica) (Fig. S17). The local atomic-scale chemical environment on the crystal basal plane could also influence ion hydration, but such study remains beyond the capability of state-of-the-art STEM.

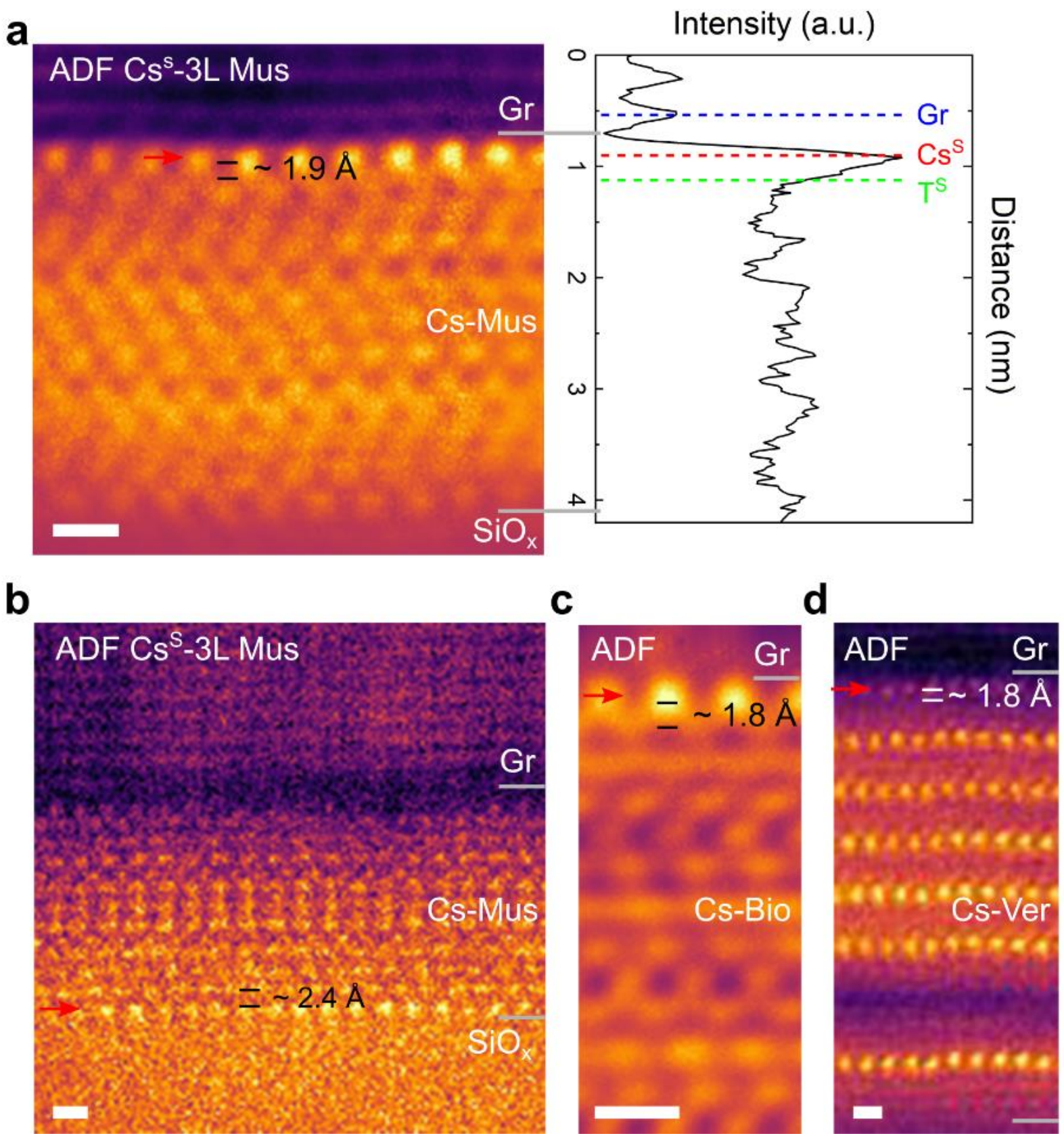

**Supplementary Figure S17 | Imaging adsorbed surface ions in cross-sectional ADF-STEM images**. **a**, Left panel, ADF-STEM image of a Cs-3L muscovite flake (graphite above and silica substrate below) with $Cs^+$ ions visible only at the interface between muscovite and Gr. Right panel, Corresponding line intensity profile showing that the $Cs^+$ peak is located at the top surface of the muscovite region with a $Cs^S$-Gr distance of ~ 4 Å. **b,** Raw data for the Cs-3L muscovite shown in Fig. 4b, main text, where $Cs^S$ surface ions are observed at the bottom surface of the muscovite (at the interface with silica). **c,** ADF-STEM image showing the top surface of a Cs-bulk biotite (interface with graphite). $Cs^+$ is adsorbed on the top surface ion sites. **d,** ADF-STEM image of a Cs-8L vermiculite. $Cs^+$ occupies the top surface sites. The red arrow indicates the $Cs^+$ ion dominated surface layer. Grey lines indicate approximate position of interfaces. Scale bars, 5 Å.

To analyse the in-plane location of exchanged ions with respect to the underlying aluminosilicate lattice we have simultaneously acquired ADF and annular bright field (ABF) STEM plan view images. Fig. S18 shows an image of a $Cs^+$ exchanged muscovite monolayer flake (exchange time 24 hours). The positions of $Cs^+$ ions are visible as bright dots in the ADF image (marked with red circles in Fig. S18c) and could be related to the aluminosilicate support (Fig. S18d). These images are enabled by a novel imaging protocol. Two sets of ABF-ADF image pairs were acquired simultaneously with a 2nm defocus difference (Fig. S19). This allowed for optimal focus, $\Delta f$, for both surface adsorbed $Cs^+$ ions (ADF $\Delta f$ = 0 nm) and the aluminosilicate lattice (ABF, $\Delta f$ = -2 nm). The best results for capturing the original $Cs^+$ sites were obtained when the $\Delta f$ = 0 nm image was acquired first (Fig. S19c). Inspection of a second ADF image (Fig. S19d) reveals that the $Cs^+$ surface ions have become significantly more disordered, preventing the identification of the original surface sites. Note however that this is not due to the loss of crystallinity in the sample – indeed, the corresponding ABF STEM image still shows a well-preserved crystal structure of the underlying aluminosilicate lattice and the native $K^+$ ions (Fig. S18b, Fig. S19f). Instead, this disorder is due to beam-induced $Cs^+$ ion displacement.

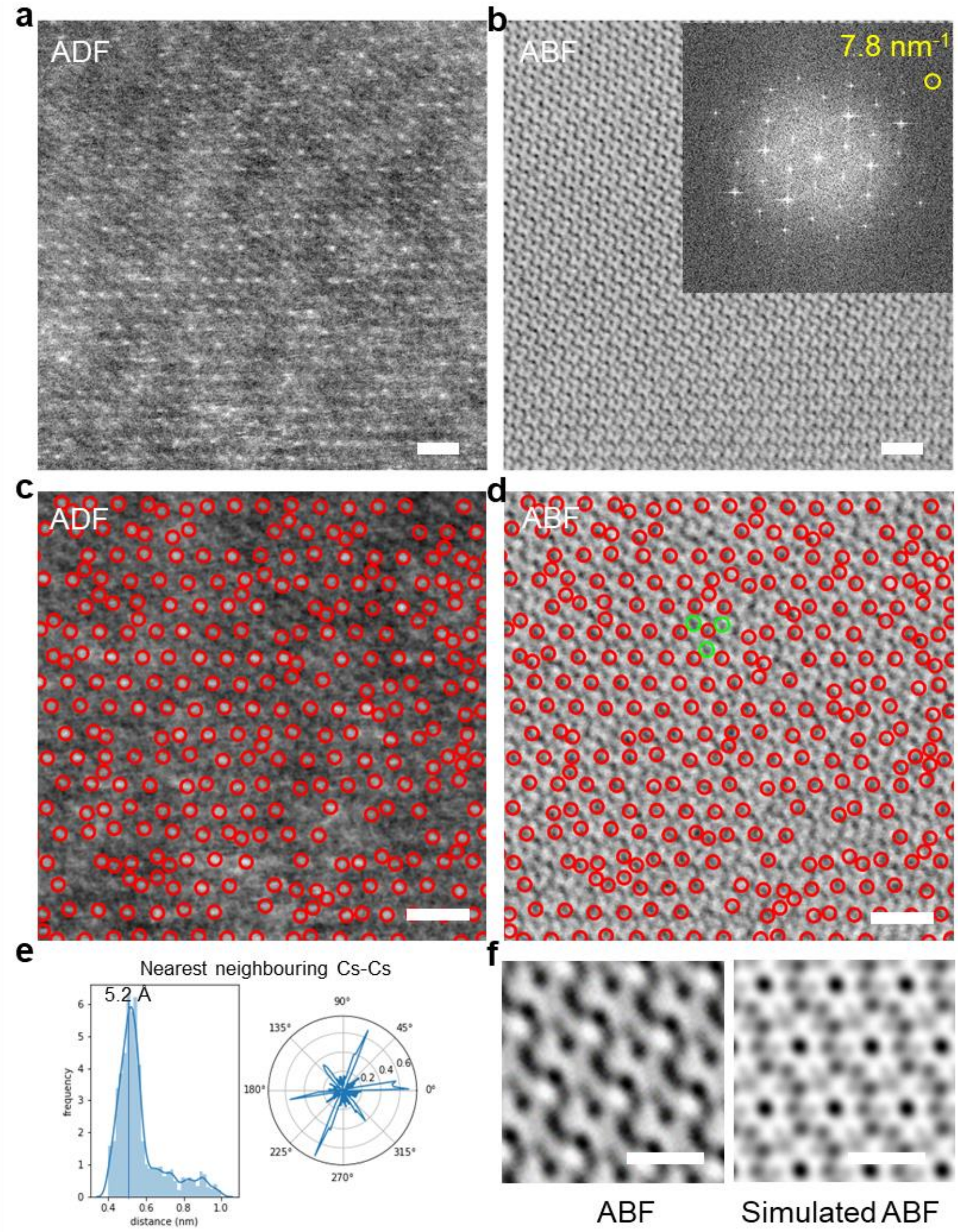

**Supplementary Figure S18|Local ordering of surface adsorbed $Cs^+$ ions on monolayer muscovite**. **a,** ADF-STEM plan view image of monolayer muscovite. $Cs^+$ ions are visible as bright dots. The slight contrast change in the top left corner of the image is due to a change in specimen height. See also Fig. S19. **b,** ABF-STEM image of the same area imaged in **a**. with the corresponding FFT (upper right inset) showing an information resolution of ~0.13 nm (i.e., 1/7.8 nm), which indicates the sample remains highly crystalline. **c,** ADF image with $Cs^+$ ion positions marked with red circles. **d**, Corresponding ABF-STEM image with Cs ion positions identified from the ADF image marked by red circles. Green circles mark the position of selected $K^+$ ions. **e,** Histogram of nearest neighbour Cs-Cs interion distance. Right panel shows the relative orientations between the neighbouring $Cs^+$ ions, demonstrating a quasi-hexagonal symmetry and local ordering of adsorbed ions aligned to the underlying aluminosilicate lattice. **f,** Left, magnified averaged experimental ABF-STEM image. Right, simulated ABF-STEM image for monolayer muscovite ($K^+$-TOT) without considering the surface $Cs^+$. **a-d**, Scale bars, 1 nm. **f**, Scale bars, 5 Å.

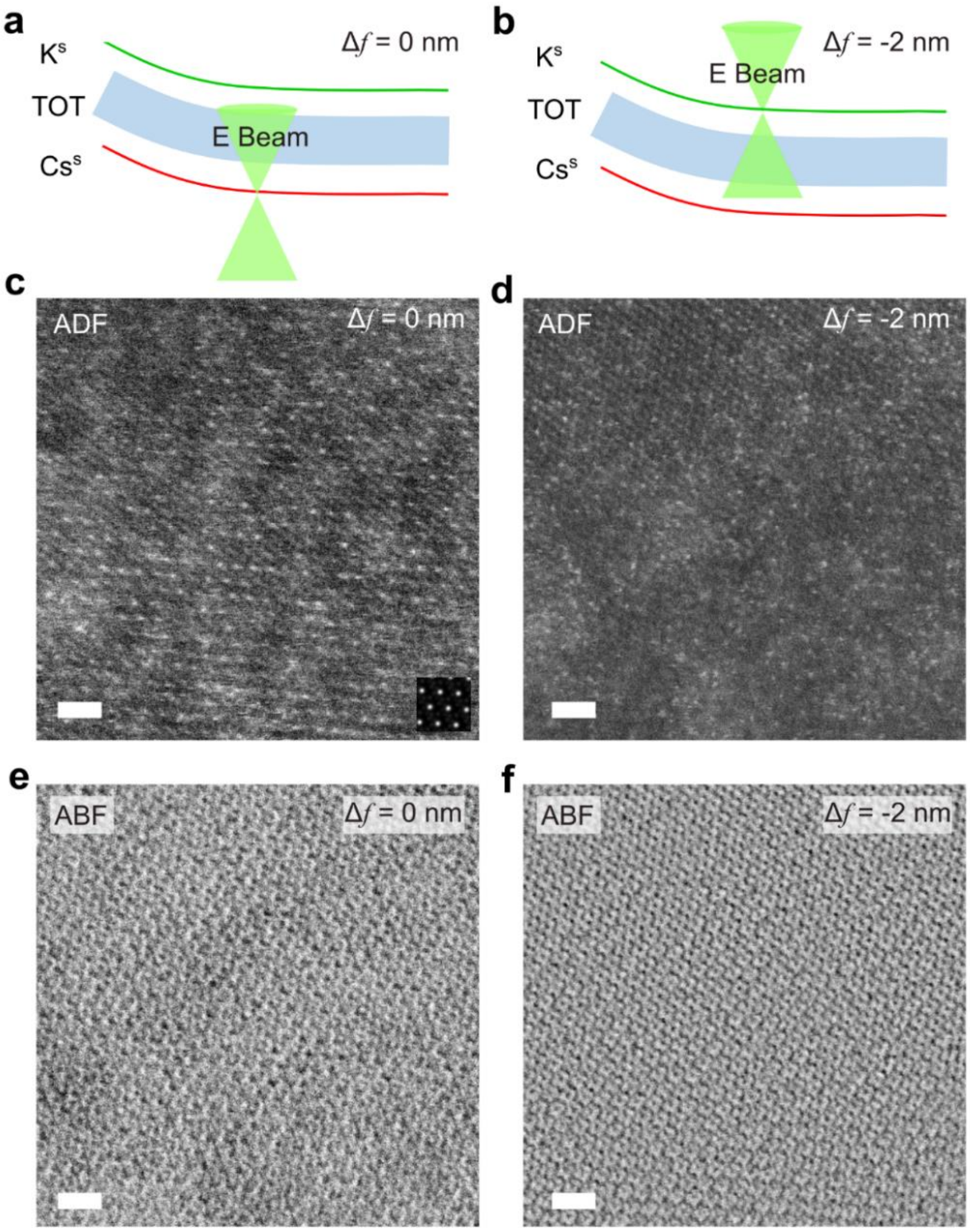

**Supplementary Figure S19|STEM Imaging protocol required to image individual adsorbed $Cs^+$ ions on monolayer muscovite**. **a, b,** Schematics showing proposed STEM focus positions for imaging surface ions on the plan-view sample shown in Fig. 4, main text. where probe focused at the bottom $Cs^+$ ion layer ($\Delta f$ = 0 nm) and the upper $K^+$ ion layer ($\Delta f$ = -2 nm). The effect of local tilt or crystal bending on defocus across the field of view is also illustrated. **c, d,** Two consecutively acquired ADF images (**c** before **d**) taken from the same area with different focus. $Cs^+$ ions are visible as bright dots and dominate the ADF contrast, with simulated ADF image ($K^+$-TOT-$Cs^+$) shown in the inset of **c**. The loss of visible $Cs^+$ ions in the top left corner of **c** is assigned to local mistilt. $Cs^+$ ions are found to be arranged in an ordered manner in **c**, while are disordered after one frame of beam exposure, as shown in **d**. **e, f,** Corresponding simultaneously acquired ABF images for **c** and **d** respectively (**e** before **f**). Scale bars, 1 nm.

For the ordered arrangement of $Cs^+$ ions (first acquisition in Fig. S18c, S19c), a statistical analysis was conducted in order to understand their surface distribution. Fig. S18e shows the calculated histogram from the nearest neighbour Cs-Cs distances as a function of relative orientation. This analysis revealed that $Cs^+$ ions form a semi-ordered hexagonal lattice with a mean inter-ion separation of ~5.2 Å. To identify the location of the $Cs^+$ ions with respect to the aluminosilicate support and confirm the expected monolayer structure, we compared the experimental images to simulations. This revealed that 228 out of 300 $Cs^+$ positions (76%) can be identified as Type II binding sites, i.e. adsorption on the vertex between three oxygen atoms in ditrigonal rings. On the other hand, nearly all $K^+$ ions were found to adsorb at the centre of the ditrigonal rings (Type I sites, see Fig. 4 in the main text). The location of selected $K^+$ ions are marked by the green circles in Fig. S18d. Only a few are indicated for clarity, but nearly all K ions were found to be located in Type I sites.

## 12. Theoretical analysis

### 12.1 Long-range forces

The expandability of the clay interlayer space is governed by a competition between short-range and long-range forces[28-30]. The former arises from Coulomb repulsion between the negatively charged aluminosilicate layers and is balanced by the interlayer cations. The attractive long-range forces arise from van der Waals (vdW) interaction between layers. For any two planes, labelled $i$ and $j$, this force is given by $F_{i,j} = A(6\pi d_{i,j}^3)^{-1}$, with $A$ the Hamaker constant and $d_{i,j}$ the space between the layers. The total attractive force on plane $j$, resulting from attraction from the other $N$-1 planes in the crystal is given by: $F_j^{(N)} = \sum_{i\neq j} F_{i,j}$. The total long-range interaction on this plane therefore depends on the number of layers in the crystal, so increasing $N$, increases $F_j^{(N)}$. However, neighbouring planes (with shorter $d_{i,j}$) exert a stronger attractive force. For this reason, $F_j^{(N)}$ converges to a limiting value with increasing $N$.

To investigate if long-range forces could account for the large interlayer expansion observed in atomically thin clays, we calculated the vdW forces on a surface plane ($F_1^{(N)}$) for clay thickness ranging from $N$ = 2-16 layers. In this calculation, we used $A$ = 1 zJ, as reported previously[29,30] and $d_{i,j}$ was extracted from our liquid AFM data (Fig. S16). We find that the attractive vdW force increases rapidly with $N$. For $N$ = 2,3 we found $F_1^{(2)} = 0.023 \times 10^4$ N m$^{-2}$ and $F_1^{(3)} = 0.19 \times 10^4$ N m$^{-2}$. However, for $N$ = 7, we find $F_1^{(7)} = 1.49 \times 10^4$ N m$^{-2}$, which is ≈60 times larger than $F_1^{(2)}$. Beyond the few-layer limit, we find that more than doubling the number of layers to $N$ = 16, only increases the force ≈1.4 times ($F_1^{(16)} = 2.18 \times 10^4$ N m$^{-2}$). This strong layer-number dependence is consistent with our STEM experimental observations. It supports the hypothesis that long-range forces are the cause for the enhanced interlayer expandability and hence the fast ion exchange.

### 12.2 Surface binding sites

Density functional theory (DFT) calculations were performed using the projector augmented wave (PAW) method implemented in the Vienna *ab-initio* Simulation Package (VASP)[31-33]. The electron exchange and correlation are described by adopting the Perdew-Burke-Ernzerhof (PBE) form of the generalized gradient approximation (GGA)[34]. The vdW force – important for layered materials – was introduced by using DFT-D2 method of Grimme[32]. The following parameters were used in our calculations. The kinetic energy cut-off of the plane-wave basis set was 500 eV in all calculations. Convergence was assumed if the energy difference between sequential steps in the iterations was below $10^{-5}$ eV. Gaussian smearing of 0.05 eV was used.

DFT was used to study the adsorption site of ions ($K^+$ or $Cs^+$) on the basal surface plane of thin muscovite. The aim is to understand the different binding sites found for $K^+$ and $Cs^+$ in the experimental data (Fig. 4c-f). To that end, the ions without solvation shells were placed at different sites on the basal plane (see Fig. S20b) and the structure was allowed to relax. We found that both ions adsorbed at the centre of the hexagonal rings, regardless of their initial position. For $K^+$ ions, this finding is consistent with our STEM images. For $Cs^+$, this lack of agreement with the simple theoretical model suggests that the situation is more complex and that the hydration shell of $Cs^+$ should play a role in its site adsorption to the aluminosilicate backbone. As mentioned in Section 11, $Cs^+$ ions on micas' surfaces can form various hydration states[18,27]. Such differences in hydration states should lead to different lateral binding sites for surface ions in the lattice[18,27]. We speculate that on average $Cs^+$ remains in the adsorption site adopted during the exchange process in solution even when the surface is being dehydrated under the high vacuum inside the microscope. Alternatively, some of these water molecules may remain associated with the Cs ions, retaining the hydration shell even in vacuum (similar to the hydrated interlayers observed in cross-sectional samples, e.g. Fig. S3h, Fig. S4, Fig. S5g). In either case, we note that even with the cutting-edge STEM analytical techniques we employed here, it is impossible to study the atomic-scale details of the hydration status for single ions in mica (clay) samples.

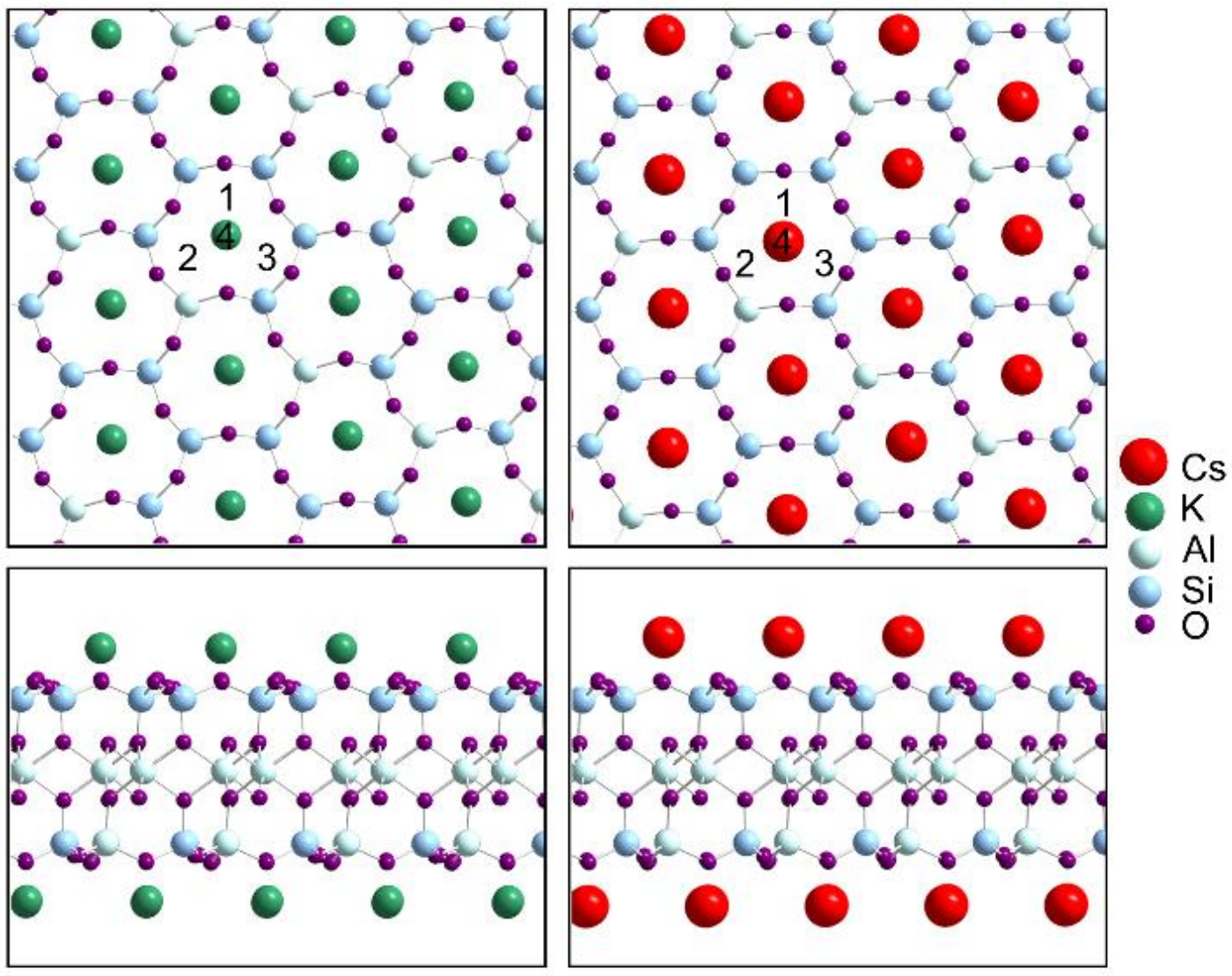

**Figure S20| DFT calculation of energies involved in the ion exchange process.** Top left, plan view DFT relaxed atomic model of the muscovite basal plane with $K^+$ ions adsorbed (without considering hydration shell). The adsorption sites at which the adsorption energy of ions was calculated is marked with black numbers labelled 1, 2 and 3 (Type II), 4 (Type I). Bottom left, cross-sectional projection of top left panel. Top and bottom right, corresponding images for $Cs^+$ exchanged muscovite.

### 12.3 Ion superlattices

To understand the mechanism behind the formation of periodic $Cs^+$ ion clusters twisted biotite micas, we modelled the potential energy landscape in their interlayer space. Each of the biotite layers was modelled using a periodic hexagonal potential (same period as in the mica lattice) and these were then superposed at a twist angle $\vartheta$ ($V(\vartheta)+V(0^\circ)$). Fig. S21 shows our results for twist angles of $\vartheta = 9.5^\circ$ and $\vartheta = 2.6^\circ$. We found that the superposed potential consists of minima (blue dots in Fig. S21c,g) at the AA stacked high-symmetry areas of the moiré unit cell, which cluster in groups of 7 and 127 for $\vartheta = 9.5^\circ$ and $\vartheta = 2.6^\circ$, respectively. Outside of these clusters, the potential becomes repulsive (red and orange areas in Fig. S21a,b,e,f). Note that the number of minima per cluster is in good agreement with the number of $Cs^+$ ions per island observed in the experiment (~7 and ~130 for $\vartheta$ = 9.5° and $\vartheta$ = 2.6°, respectively). This indicates that the energy necessary to trap ions is 77% of the single-layer potential minima. Furthermore, we find that these clusters occur periodically and with the same periodicity found experimentally for the $Cs^+$ superlattices. These results therefore validate the use of this model, which attributes the formation of $Cs^+$ superlattice to moiré potential 'wells'.

The model also helps to explain the large enhancement in ion exchange speed found in twisted biotite compared to the pristine aligned interlayers. Fig. S21d,h show that in a twisted system (red curves), for sites outside the high-symmetry AA region of the moiré unit cell, the interlayer potential energy is higher (more positive/repulsive) than that in the aligned bilayer system (green curves). This results in a weaker interlayer binding in twisted materials compared to that in a pristine untwisted one, which in turn enhances interlayer expandability and ion diffusion.

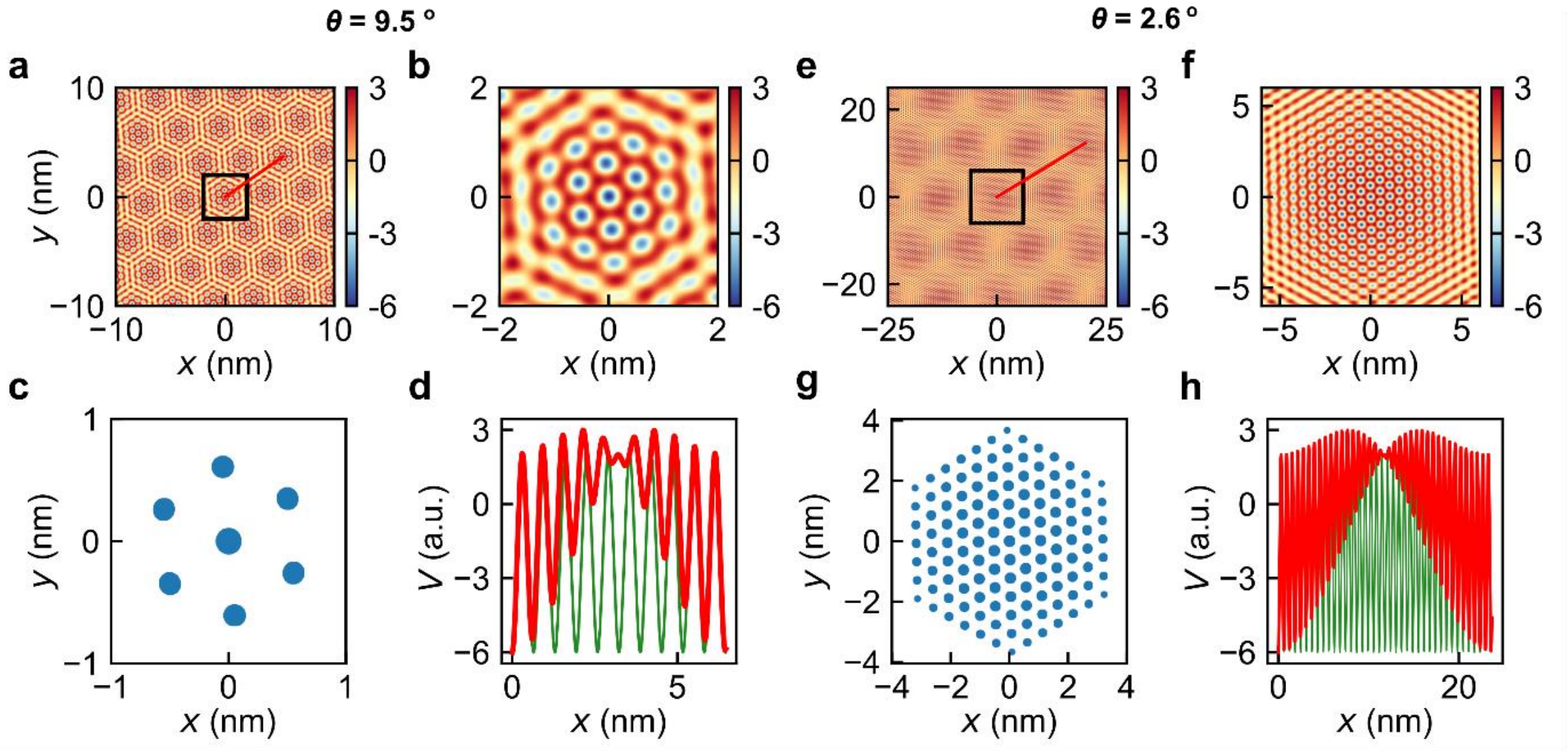

**Figure S21| Model of interlayer potential energy landscape in twisted biotite. a,** Potential energy map resulting from the superposition of two hexagonal periodic potentials with a twist angle of $\vartheta$ = 9.5°, each with a period of 0.53 nm. The resulting moiré potential has a period of 3.2 nm. Energy scale in arbitrary units. **b**, Magnified region of the potential energy map from the high-symmetry area marked by the black square in panel **a**. **c**, Potential energy map with blue dots marking the areas where the energy minima is 77% as deep or deeper than the minima of the pristine single-layer potential. **d**, Red line is the potential energy profile taken along the red line in panel **a** (drawn from one AA site to a neighbouring one) compared to the equivalent line profile taken from a perfectly aligned (untwisted) bilayer system (green line). **e,f,g,h**, Corresponding results for a smaller twist angle, $\vartheta$ = 2.6°.